\documentclass[draft]{agujournal2019}
\usepackage{url} 
\usepackage{lineno}
\draftfalse

\journalname{JGR: Atmospheres}

\begin{document}

%
%


\title{CloudSat-inferred vertical structure of precipitation over the Antarctic continent}

%
%




\authors{F. Lemonnier\affil{1}, J-B. Madeleine\affil{1}, C. Claud\affil{1}, C. Palerme\affil{2}, C. Genthon\affil{1}, T. L'Ecuyer\affil{3} and N. B. Wood\affil{3}}

\affil{1}{Sorbonne Universit\'e, \'Ecole normale sup\'erieure, PSL Research University, \'Ecole polytechnique, CNRS, Laboratoire de M\'et\'eorologie dynamique, LMD/IPSL, F-75005 Paris, France}
\affil{2}{Development Centre for Weather Forecasting, Norwegian Meteorological Institute, Oslo, Norway}
\affil{3}{Department of Atmospheric and Oceanic Sciences, University of Wisconsin-Madison, Madison, Wisconsin, USA}





\correspondingauthor{Florentin Lemonnier}{florentin.lemonnier@lmd.jussieu.fr}




\begin{keypoints}
\item Regridding of the CloudSat observations reveals the 3D structure of precipitation over Antarctica 
\item Distributions vary substantially by type of region: ice sheet, peninsula or ice shelves
\item Precipitation rate - temperature relationship is explained by large-scale orographic precipitation
\end{keypoints}

%
%


\begin{abstract}

Current global warming is causing significant changes in snowfall in polar regions, directly impacting the mass balance of the ice caps. The only water supply in Antarctica, precipitation, is poorly estimated from surface measurements. The onboard cloud-profiling radar of the CloudSat satellite provided the first real opportunity to estimate precipitation at continental scale. Based on CloudSat observations, we propose to explore the vertical structure of precipitation in Antarctica over the 2007-2010 period. A first division of this dataset following a topographical approach (continent versus peripheral regions, with a 2250~m topographical criterion) shows a high precipitation rate (275~mm.yr$^{-1}$ at 1200~meters above ground level) with low relative seasonal variation ($\pm$11$\%$) over the peripheral areas. Over the plateau, the precipitation rate is low (34~mm.yr$^{-1}$  at 1200~m.a.g.l.) with a much larger relative seasonal variation ($\pm$143$\%$). A second study that follows a geographical division highlights the average vertical structure of precipitation and temperature depending on the regions and their interactions with topography. In particular, over ice-shelves, we see a strong dependence of the distribution of precipitation on the sea-ice coverage. Finally, the relationship between precipitation and temperature is analyzed and compared with a simple analytical relationship. This study highlights that precipitation is largely dependent on the advection of air masses along the topographic slopes with an average vertical wind of 0.02~m.s$^{-1}$. This provides new diagnostics to evaluate climate models with a three-dimensional approach of the atmospheric structure of precipitation.

\end{abstract}

\section{Introduction}

Predicting the mass balance of ice sheets is a major challenge in the context of sea level rise. This surface mass balance depends on the relative magnitudes of precipitation versus sublimation/evaporation, meltwater run-off and blowing snow \cite{eisen2008ground}. On an unfriendly continent for field observations, satellites are keys for observing precipitation. Based on these spatial remote sensing observations, recent studies observe an increasing ice loss in West Antarctica due to the acceleration of glacier and basal melting \cite{shepherd2018mass,shepherd2012reconciled, pritchard2012antarctic}. This does not seem to be restricted to the West alone, and \citeA{rignot2019four} shows that ice loss is more significant than its gain at the scale of the entire continent.

According to the climate models of the 5$^{th}$ Climate Model Intercomparison Project, precipitation over Antarctic would increase from 5.5$\%$ to 24.5$\%$ between 1986-2005 and 2080-2099. This increase would have a significant impact on sea level \cite{church2013sea}. However, the present-day averaged Antarctic snowfall rates predicted by these models range from 158 to 354 mm.yr$^{-1}$ while the first model-independent climatology of current Antarctic precipitation yields a value of 172 mm.yr$^{-1}$ over the 2007 -- 2010 period \cite{palerme2017evaluation}. The latter value was recently re-evaluated at 160 mm/year \cite{palerme2018groundclutter} over the 2007 -- 2010 period. It is worth noting that \citeA{milani2018cloudsat} have also produced a climatology that also includes the Southern Ocean.

Apart from these surface precipitation estimates, the vertical structure of precipitation was poorly known until recently, but work by \cite{grazioli2017katabatic} compared vertical year-averaged profile of precipitation observed with ground radars with simulated snowfall profiles at Dumont d'Urville station. It showed surface rates in good agreement with climate models but significant biases in the vertical structure of precipitation. Using long periods of ground-based observations to characterize local snowfall events showed also a significant amount of virgas in precipitation events \cite{maahn2014,duranalarcon2018VPR}. A poor representation of the vertical structure of precipitation reveals deficiencies of process representation in models including microphysical processes.

The thermal structure of the atmosphere over Antarctica, which is deeply involved in the origin of precipitation, is unique. The center of the continent being a cold pole, the structure of the Antarctic circumpolar flow is of baroclinical type. Low pressure systems are thus generated by horizontal temperature gradients in the troposphere and grow through baroclinic instabilities. Low pressure systems bring strong winds to the coastal areas when pressure gradients increase as polar lows over the ocean are moving southward and encounter areas of higher pressure, such as the semi-permanent continental anticyclone \cite{bromwich1998meteorology,king2007antarctic,van2007heat}. These winds are the moisture vectors that control large scale precipitation processes. Indeed, winds move air masses against the Antarctic continent and lead to encounters of moist and dry, hot and cold air masses, thereby creating ideal conditions for the formation of precipitation. Over the vertical axis, precipitation undergoes various changes before it reaches the ground. The origin of precipitation is described by cloud microphysics. Microphysical processes such as ice crystal nucleation are initiated in clouds, grow by diffusion at the expense of the supercooled droplets or by collision with other crystals, and fall to the ground \cite{findeisen2015colloidal,dye1974mechanism}. During its downfall, precipitation can reach drier air masses and sublimate, which can significantly reduce the precipitation rates. On the contrary, precipitation can reach saturated air masses, interact with potential clouds at lower altitude, aggregate and increase its mass flow. It is precisely all these processes and vertical evolution that define surface precipitation. 

Precipitation in Antarctica is challenging to study, mostly due to geographic characteristics. Indeed, ground-based measurements are sparse and challenging in Antarctica and the size of this continent does not allow to cover and study the whole distribution, frequency and rate of precipitation. In coastal areas, it is influenced by synoptic conditions such as oceanic fronts \cite{bromwich1988snowfall} and it is difficult to distinguish between precipitation and blowing snow caused by strong katabatic winds. Furthermore, re-sublimation processes of the precipitation in the lower layer of the atmosphere have been observed at Dumont d'Urville with a micro-rain radar, and result in a decrease in the precipitation rates at the surface \cite{grazioli2017katabatic}. At Princess Elisabeth and Dumont d'Urville, the study of \citeA{duranalarcon2018VPR} showed that in about a third of precipitation event cases, they are virgas that do not reach the surface. There can also be advection of very large amounts of moisture, providing a high annual precipitation contribution over the continent \cite{gorodetskaya2014role}. According to ground observations over peripheral areas and plateau, the ice crystals that are at the origin of the precipitation are thinner than 100 $\mu$m and aggregate to constitute coarser particles \cite{lachlan2001observations,lawson2006microphysical}. A recent study shows that the occurrence of ice crystals at Dumont d'Urville is similar to that of aggregates at low altitude ($<$500 m above the surface) \cite{grazioli2017measurements}. Over the continental plateau ($>$2250 m), most of the precipitation is driven by a few fronts while the remaining annual precipitation rate is in the form of ``Diamond Dust'' (thin ice crystals) under clear-sky conditions \cite{bromwich1988snowfall,fujita2006stable,turner2019dominant}.

The first real opportunity to assess precipitation in polar regions from a spaceborne radar platform appeared with the cloud-profiling radar (CPR) on CloudSat satellite \cite{stephens2008controls,liu2008cloud}. CloudSat provided day- and night-time solid precipitation observations from August 2006 to April 2011, which led to the first multi-year, model-independent climatology of Antarctic precipitation, with an average of 171 mm w.e year$^{-1}$ over the Antarctic ice sheet, north of 82$^{o}$S \cite{palerme2014much}. The first hundreds of meters over the surface are contaminated by radar wave reflections from the surface, which is called ground-clutter. A study of \citeA{maahn2014} investigated whether a thinner ground clutter layer could increase the quality of the precipitation measured at Princess Elisabeth, but the data are really usable only from 1200 m.a.g.l. The climatology of \citeA{palerme2014much} is based on the 2C-SNOW-PROFILE product \cite{wood2011estimation},, which provides, on average, an uncertainty on single snowfall retrievals ranging between 1.5 and 2.5 times the snowfall rate \cite{wood2011estimation,palerme2014much}. However, recent studies based on comparisons with ground-based radars reassessed this range of uncertainty for a few snowfall events at two Antarctic stations \cite{lemonnier2018evaluation,souverijns2018cloudsatgrid}, and provided uncertainties ranging from 13 to 22$\%$ for the CloudSat snowfall observations. \citeA{palerme2018groundclutter} have identified some CloudSat observations likely contaminated by the ground clutter over mountainous areas, and have produced a new climatology with a mean snowfall rate of 160 mm.yr$^{-1}$.

In this study, we computed the first 3-D multi-year climatology of Antarctic precipitation north of 82$^\circ$S from spaceborne remote sensing observations (section \ref{sect:data_methods}). As a first step, we characterize the general horizontal and vertical structures of the precipitation averaged in time over the continent (section \ref{sect:3D_structure}). Then, a comparison between CloudSat snowfall retrievals and concurrent 3D meteorological parameters obtained from the European Center for Medium-Range Weather Forecast (ECMWF) operational weather analysis at each vertical level is conducted in section \ref{sect:scatterplots_precip}. This allows us to determine which physical or dynamical process predominates on the high continental plateau and on the peripheral areas. At last, we propose a new way of evaluating modelled precipitation using this new dataset, and especially its vertical structure, in atmospheric models such as global GCMs or Antarctic regional models (section \ref{sect:conclusion}).


\section{Data and method}
\label{sect:data_methods}

Here, the climatology produced by \citeA{palerme2018groundclutter} is extended to all vertical levels up to 10000~m. The first 4 bins above the surface are excluded due to potential ground clutter contamination \cite{wood2011estimation}. We used the fifth release of 2C-SNOW-PROFILE CloudSat product \cite{wood2011estimation,wood2014estimating} that gives an estimate of the liquid-equivalent snowfall rate and its Snow Retrieval Statue (SRS) which indicates if some errors occurred in the retrieval process. The retrieval of this product is based on a relationship between reflectivity and snowfall that takes into account a priori estimates of snow particle size distribution and microphysical and scattering properties at each vertical bin \cite{rodgers2000inverse}. Then this product is averaged within each cell of a grid over the Antarctic continent from about 1200~m.a.g.l. (meters above ground level) up to 10000~m.a.g.l. (5$^{th}$ to 45$^{th}$ bins above the surface) with a 240~m vertical resolution. The same extraction process is applied to the associated ECMWF-AUX operational weather analysis temperatures, humidities and pressures in order to obtain a climatology based on the same sampling.

The low temporal sampling of CloudSat is a source of uncertainty. Following \citeA{souverijns2018cloudsatgrid}, this new CloudSat climatology is processed over a grid of 1$^\circ$ of latitude by 2$^\circ$ of longitude. This grid is able to accurately represent a snowfall climatology and is the best balance between satellite omissions and commissions, according to data from three ground radars. 

2C-SNOW-PROFILE provides a snowfall uncertainty between 1.5 and 2.5 times the solid precipitation rate for a single measurement. The average over the whole continent over the 2007--2010 period reduces this uncertainty but some uncertainty remains due to systematic errors in algorithm assumptions that are difficult to assess \cite{palerme2014much}. \citeA{lemonnier2018evaluation} reassessed this range of uncertainties by a short-time (a few seconds) and short-spatial scale comparison of CloudSat measurements with 24 GHz Micro Rain ground radars at different locations in East Antarctica (a coastal and a high altitude station) with two different kinds of weather conditions, leading to new values of [-13$\%$, +22$\%$] and providing confidence over the full height of the CloudSat retrievals.

The temperature associated with each precipitation profile is obtained from ECMWF-AUX operational weather analysis. The ECMWF-AUX data set is an intermediate product that contains the set of ECMWF state variable data interpolated to each CloudSat cloud profiling radar bin. Besides the snowfall rate climatology computed on a continental grid at each vertical level of the satellite, the pre-gridded dataset of the 2C-SNOW-PROFILE product (snowfall retrieval and geolocation fields) as well as the ECMWF-AUX product were extracted in order to evaluate the relationship between precipitation rates and temperatures (see section \ref{sect:scatterplots_precip}).

\section{General structure of Antarctic precipitation}
\label{sect:3D_structure}

\begin{figure*} [!h]
	\centering
	\includegraphics[width=1\linewidth]{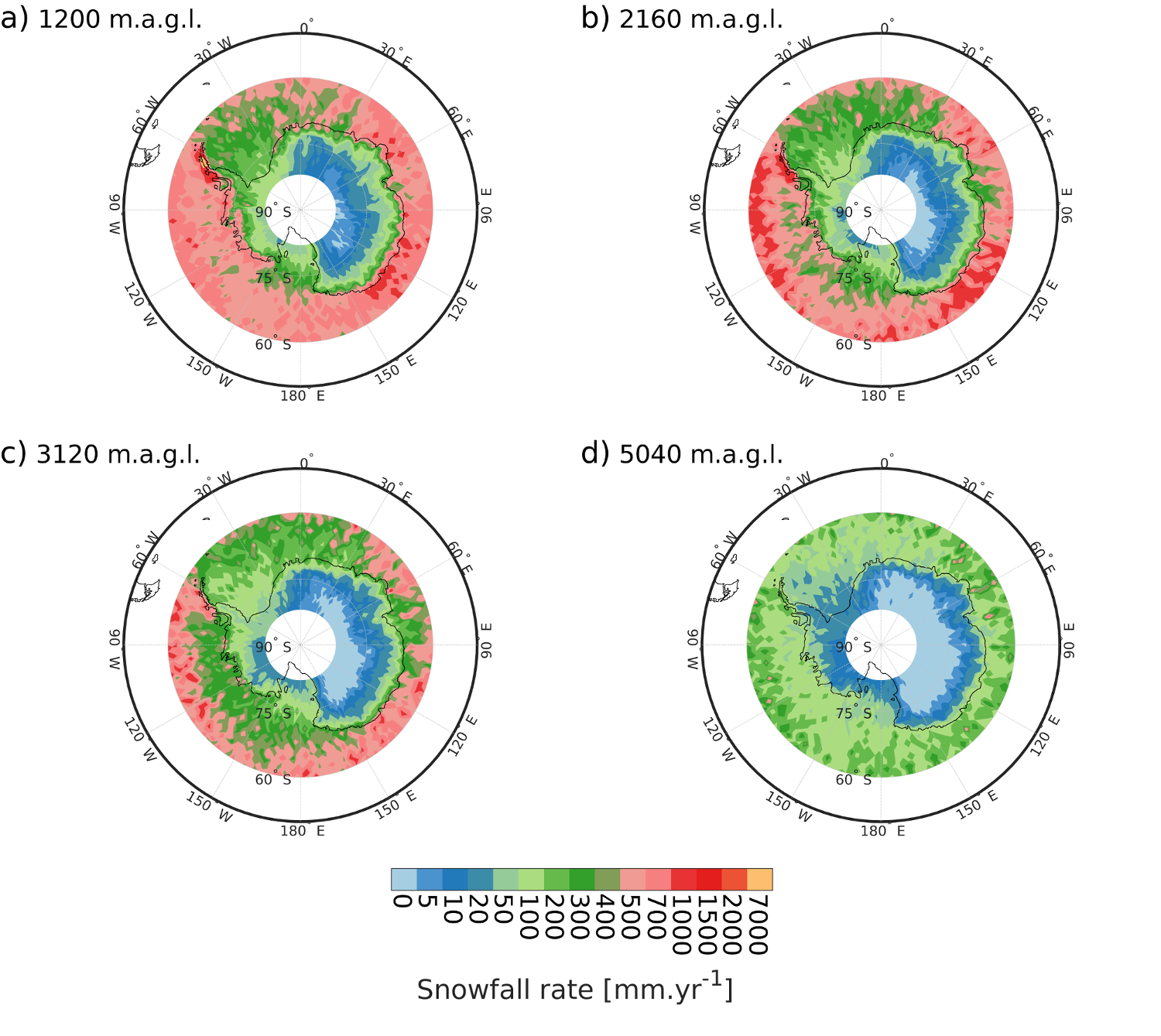}
	\caption{\label{Cartes}
	Mean annual snowfall rate (mm water equivalent / year -- mm.yr$^{-1}$) from the 2C-SNOW-PROFILE product over the 2007--2010 period \textbf{a)} at 1200 m.a.g.l. (5$^{th}$ bin); \textbf{b)} at 2160 m.a.g.l. (8$^{th}$ bin); \textbf{c)} at 3120 m.a.g.l. (12$^{th}$ bin) and \textbf{d)} at 5040 m.a.g.l. (20$^{th}$ bin).
	}
	\end{figure*}

Figure \ref{Cartes} shows precipitation maps over Antarctica at four different CloudSat vertical levels. As highlighted in previous studies \cite{palerme2014much,milani2018cloudsat}, there is a dichotomy in precipitation rate between the high plateau of the continent where the snowfall rate is excessively low and the coastal areas. It is also important to note extremely high precipitation rates located west of the Antarctic peninsula and over mount Vinson. This is due to high orography and moisture flux from the Southern Ocean, producing more clouds and precipitation \cite{nicolas2011climate}. Precipitation over the ice-shelves appears to be very low. According to the study of \citeA{knuth2010influence} carried out over the Ross ice-shelf, it even appears that the blowing and drifting snows prevails over the accumulation of precipitation. Thus, precipitation is regionally very variable in Antarctica. Figure \ref{Cartes} shows that precipitation decreases with altitude over the continent, which is not always the case over the ocean. Indeed, over all oceanic regions south of 60$^\circ$S except the eastern part of the peninsula and above the ice-shelves, the precipitation rate increases between the 1200 m.a.g.l. (fig. \ref{Cartes}.a) and 2160 m.a.g.l. (fig. \ref{Cartes}.b) vertical bins before decreasing further at altitude (fig. \ref{Cartes}.c and \ref{Cartes}.d). This could be explained by a local formation of snow in regions of shallow convection, as highlighted by \citeA{kulie2016shallow} and \citeA{milani2018cloudsat}, who looked at the differences between the 3$^{rd}$ and the 6$^{th}$ vertical bins.

	\begin{figure*} [!ht]
	\centering
	\includegraphics[width=1\linewidth]{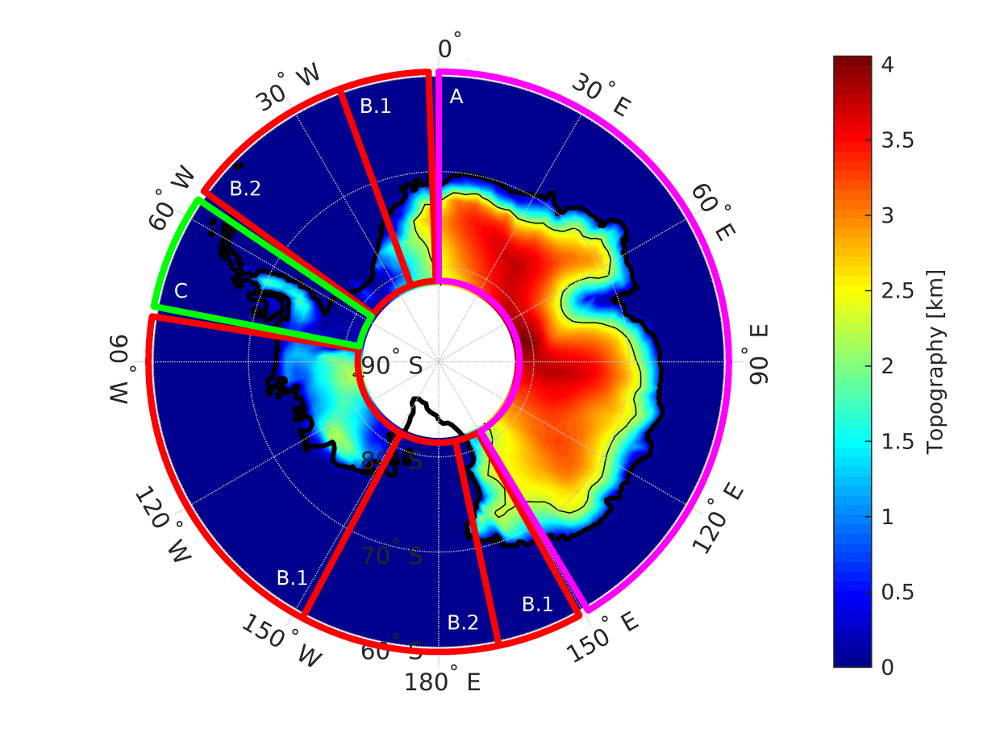}
	\caption{\label{Camembert}
	Digital Elevation Map of the Antarctic Ice Sheet (Liu et al., 2015) with four frames corresponding to the four studied areas. \textbf{A} in magenta is the East Antarctic continent; \textbf{B} in red is the West Antarctic, which has been subdivided into continental region (B.1) and ice-shelves (B.2); \textbf{C} in green is the Peninsula. Thin black solid line is the iso-altitude 2250 m and separates the coastal areas from the plateau.
	}
	\end{figure*}

The thin black solid line in fig. \ref{Camembert} delineates regions that are 2250 m above sea-level. This altitude separates peripheral areas where precipitation is mostly controlled by large scale oceanic fronts, from the plateau where snowfall is rare and depends on a few events which are large enough to bring moisture onto the Antarctic plateau.

In order to have a better understanding of the horizontal and vertical structures of snowfall, the Antarctic continent is also divided into several geographic regions. Figure \ref{Camembert} presents these different regions for the following geographic areas :
\begin{itemize}
	\item \textbf{A} The first studied area is the East Antarctic, located between 0$^\circ$ and 150$^\circ$ of longitude. This area is defined mostly as an homogeneous ice-sheet with a huge plateau at an altitude of 3000 m above sea level.
	\item \textbf{B} The second studied area is the West Antarctic, lower in altitude than the East part of the continent. This region has been subdivided in two parts, one comprising mountain ranges including the mount Vinson (B1) and the other part comprising the massive Ross and Ronne-Filchner ice-shelves (B2).
	\item \textbf{C} The third region of interest is the peninsula, which is very mountainous and where precipitation is exceptionally high.
\end{itemize}

\subsection{From coasts to high continental plateau}

\begin{figure*} [h!]
	\centering
	\includegraphics[width=0.5\linewidth]{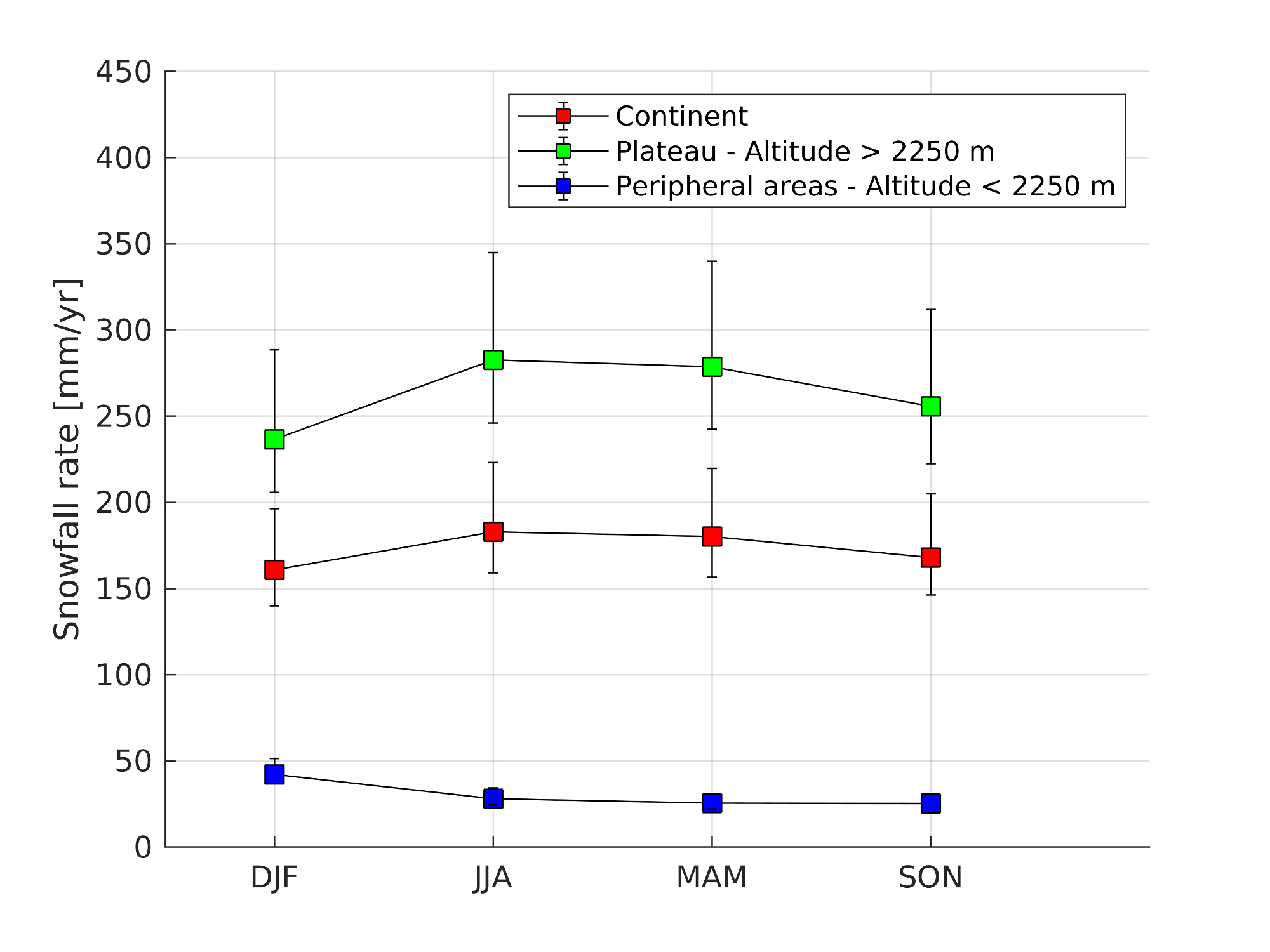}
	\caption{\label{Palerme} Seasonal variability of snowfall north of 82$^\circ$ during the period 2007-2010 for CloudSat in mm/yr at about 1200~m.a.g.l. over the entire continent in red, over the peripheral areas in green and over the high continental plateau in blue. Errorbars represent uncertainties as calculated by \citeA{lemonnier2018evaluation} and extrapolated to the entire continent.
	}
	\end{figure*}

By dividing Antarctica according to topographical information (2250~m in surface elevation, see \citeA{palerme2014much}), we distinguish two different types of climate. The peripheral areas includes the Peninsula (surface elevation $<$ 2250 m), West Antarctica and eastern coasts. The plateau (surface elevation $>$ 2250 m) includes essentially East Antarctica.
Following \citeA{palerme2017evaluation}, fig. \ref{Palerme} summarizes the seasonal evolution of the precipitation over the Antarctic continent. The seasonal variability of snowfall across the Antarctic continent is mainly influenced by snowfall in peripheral regions, with the periphery of the continent receiving the vast majority of precipitation \cite{bromwich1988snowfall,genthon2009antarctic}. This figure corresponds to the 5$^{th}$ CloudSat vertical bin at about 1200 m.a.g.l. Over the Antarctic continent and especially over peripheral areas (altitude $<$ 2250~m), the maximum snowfall rate is observed in March-April-May (MAM) then decreases until December-January-February (DJF) where the snowfall rate is minimum. Over the plateau, this variability differs with a maximum in precipitation rate over the DJF period and a minimum in June-July-August (JJA). Error bars represent the CPR measurement uncertainties given by \citeA{lemonnier2018evaluation}, obtained by a comparison of four precipitation events with ground instruments at two different locations. The seasonal evolution of precipitation has been studied for the other vertical levels of the product, and no difference in seasonal variability from one level to another is observed. As already mentioned by \citeA{palerme2017evaluation}, the seasonal evolution of mean precipitation in Antarctica is in accordance with ERA-Interim and climate models for the peripheral regions. This is not the case for precipitation above the plateau. A more recent study by \citeA{lenaerts2018signature} focusing on ground accumulation presents similar results, although it focuses on surface snow accumulation, where other processes such as surface sublimation or blowing snow are at work.

\begin{figure*} [h!]
	\centering
	\includegraphics[width=1\linewidth]{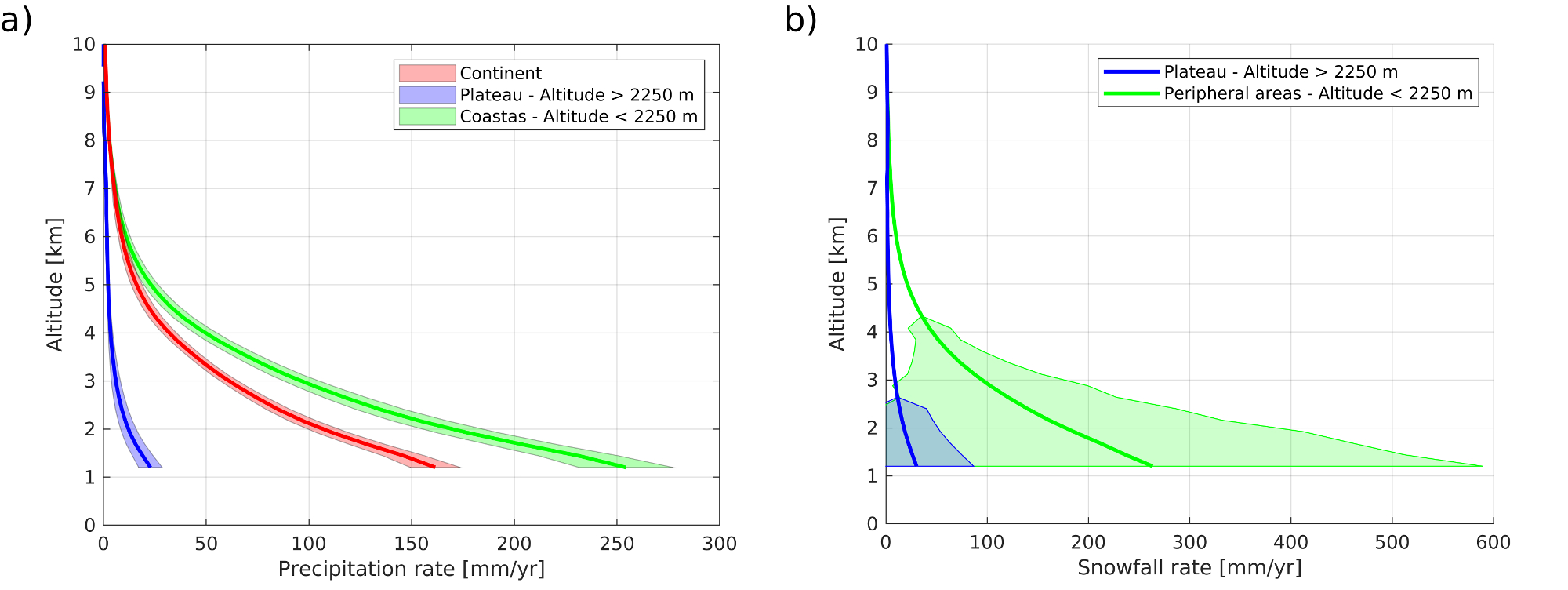}
	\caption{\label{Profiles} \textbf{a)} Averaged vertical profiles of precipitation over the 2007-2010 period of CloudSat observations in solid lines, filled areas are corresponding to the temporal $\sigma$ standard deviations
	\textbf{(1)} over the entire continent in red.
	\textbf{(2)} over the plateau in blue.
	\textbf{(3)} over the peripheral areas in green. \textbf{b)} Same results as \textbf{a)} but the filled areas correspond to the spatial $\sigma$ standard deviations \textbf{(1)} over the plateau in blue.
		\textbf{(2)} over the peripheral areas in green.
	Over the peripheral areas, the first bin value is 275 mm.yr$^{-1}$ $\pm$ 11$\%$, the precipitation rate at the first bin over the Antartic plateau is 34 mm.yr$^{-1}$ $\pm$ 143$\%$ and the precipitation for the entire continent north of 82$^\circ$ is 163 mm.yr$^{-1}$ $\pm$ 13$\%$.
	}
	\end{figure*}

Fig. \ref{Profiles} shows another aspect of the vertical structure of precipitation in Antarctica above the ground clutter layer represented by annual average precipitation profiles. Each profile is averaged over the 2007--2010 period, then area-averaged. The standard deviation on fig. \ref{Profiles}.a is calculated from the time dimension (in months) and on fig. \ref{Profiles}.b it is calculated from the gridded spatial dimension. As observed in this figure, monthly temporal variability is low over the Antarctic continent. We note that the time standard deviation is continuous over the entire precipitation profile. The spatial variability is important, whether on the plateau or over peripheral areas. This spatial standard deviation decreases very rapidly with altitude. Thus, the precipitation rates recorded at high altitudes appear to be regionally homogeneous. Precipitation consistently increases from 7000 m in altitude until 1200 m, reaching a rate of approximately 160 mm.yr$^{-1}$. At this altitude, the precipitation gradient should become negative and the profile should reach a maximum in precipitation before inverting in the lowest layers over the peripheral areas (in green), according to several studies \cite{grazioli2017katabatic,duranalarcon2018VPR}. The blue line presents the profile over the plateau. It is characterized by a very small amount of precipitation all along the profile and a relatively large dispersion : 30 mm.yr$^{-1}$ with a $\sigma$-value of almost 50 $\%$ in the lowest bin. It suggests a high variability over the plateau. In comparison with the plateau, the green line presents the peripheral profile of precipitation, with higher precipitation values and a lower variability. Fig. \ref{Profiles} confirms that precipitation in Antarctica mostly occurs over the coasts. As highlighted by \citeA{grazioli2017katabatic}, there is an inversion of the precipitation profile in the lowest layers ($<$ 1000~m) over the Dumont d'Urville station due to snow sublimation by katabatic winds. However, CloudSat does not see this low-level sublimation inversion layer because of ground clutter \cite{lemonnier2018evaluation,palerme2018groundclutter}.

\subsection{From geographical areas}

\begin{figure*}[!p]
	\centering
	\includegraphics[width=1\linewidth]{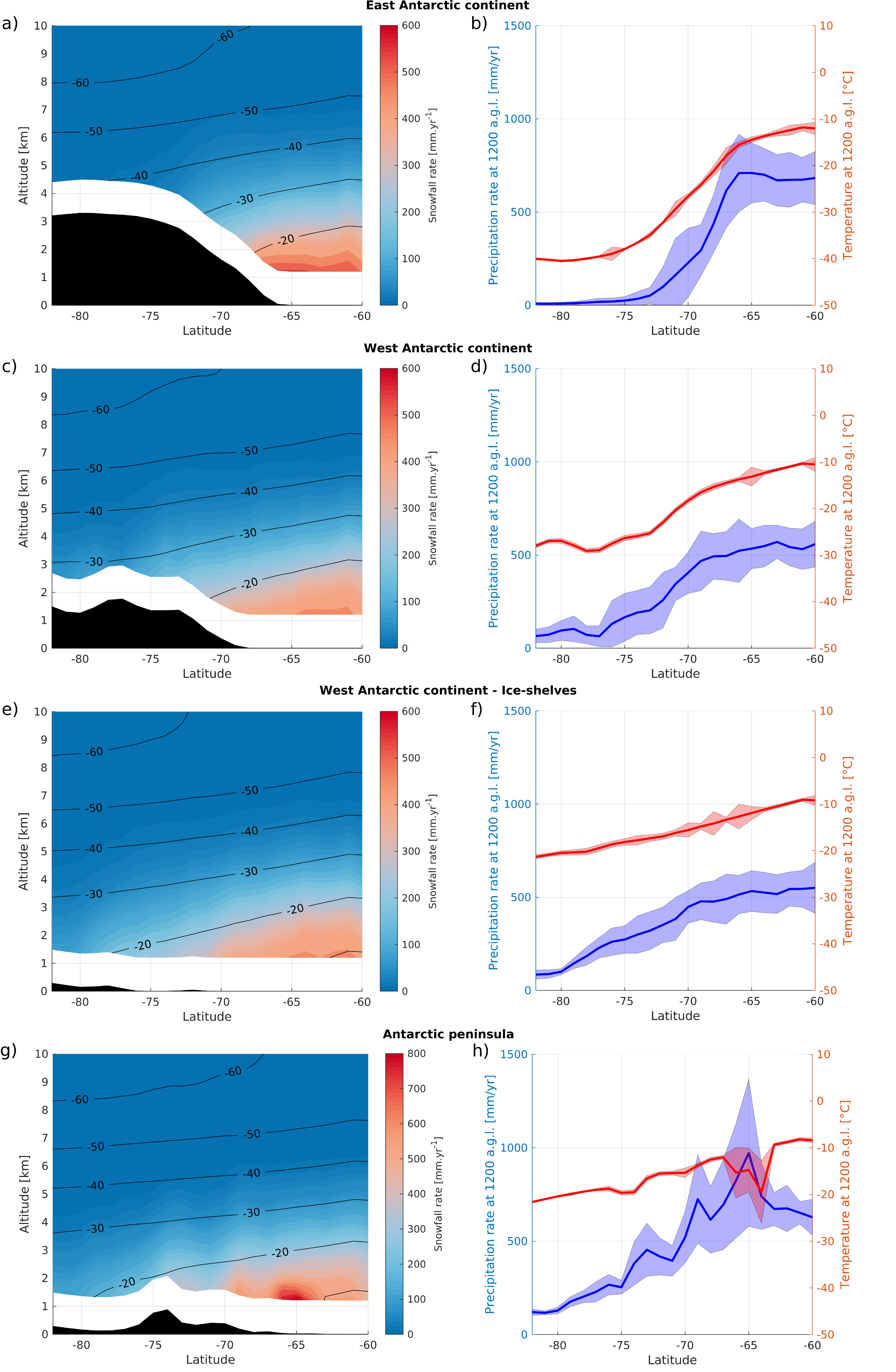}
	\caption{\label{3D_struct}
	First column presents the zonal averaged precipitation rate (shaded colors) and its corresponding atmospheric temperature (contours) of each region, as follows : \textbf{a)} for East continent (A in fig.\ref{Camembert}), \textbf{c)} for West continent (B.1 in fig.\ref{Camembert}), \textbf{e)} for West ice-shelves (B.2 in fig.\ref{Camembert}) and \textbf{g)} for Peninsula (C in fig.\ref{Camembert}). Black filled areas correspond to the zonal-average elevation of each area. Second column presents the fifth CloudSat level (1200 m.a.g.l.) of precipitation along the latitude (in blue) with its $\sigma$-standard deviation calculated over the spatial dimension. In red the associated temperature and its $\sigma$-standard deviation are represented.
	}
	\end{figure*}

The re-gridded precipitation dataset is used to evaluate seasonal variabilities as well as vertical variabilities. This study also takes into account different regions of interest with the aim of providing useful diagnostics for climate models. Precipitation in each geographical region is longitudinally averaged at each CloudSat vertical bin for the entire period of observation. The same computation is applied to the atmospheric temperature from ECMWF operational weather analysis and digital elevation model. The results are presented in the first column of Fig. \ref{3D_struct}. The baroclinic structure of temperature, with the isotherms bending down towards the pole, is visible on each latitudinally averaged regions of fig. \ref{3D_struct}. Then we present in the second column the averaged snowfall rate at the fifth CloudSat level (at 1200 m.a.g.l., same studied altitude above ground level than \citeA{palerme2014much}) in blue and its associated temperature in red over both the continent and the oceans (up to 60$^\circ$). The filled areas around the means in blue and red are the corresponding $\sigma$-standard deviations of the precipitation rate and temperature. The standard deviations are calculated over the spatial dimension for each region of interest.
	
Fig. \ref{3D_struct}.a represents precipitation structure over East Antarctica, the most homogeneous region of the Southern continent, and highlights an obstacle to the progression of precipitation flows with topography. The iso-precipitation rates are following isotherms over the ocean but are no longer parallel over the continent with a faster decrease in precipitation isolines. The precipitation rate drops from 600 mm.yr$^{-1}$ to less than 10 mm.yr$^{-1}$ along 10 degrees of latitude and 3000 meters of altitude difference. Fig. \ref{3D_struct}.b, which only considers retrievals at the first available bin over the surface, shows with more accuracy this latitudinal evolution. The precipitation rate over the plateau is very low and its associated standard deviation is relatively important (almost 100$\%$). From 75$^\circ$S and northward, at the edge of the plateau, precipitation rate increases quickly before reaching a maximum value of $\sim$600 mm.yr$^{-1}$ over the ocean at latitude 66$^\circ$S. The temperature along the slope follows a similar evolution. Over the Southern Ocean, precipitation and temperature stabilize and reveal a more homogeneous oceanic climate.

Regarding the western part of the continent, there are some similarities with the eastern continent, such as the low precipitation rate on the plateau on fig. \ref{3D_struct}.c. The average precipitation rate at low altitude is lower than in the East, at 500 mm.yr$^{-1}$. The transition from continental to oceanic precipitation rates seems to be more gradual. The precipitation rate decreases from 500 mm.yr$^{-1}$ to about 150 mm.yr$^{-1}$ at an average elevation of 1500 m without stabilizing above the plateau, probably due to local orographic effects that characterize this region. The sea-ice cover is vast in this region of Antarctica, although the seasonal variability of its area is high, this precipitation pattern may be a persistent signal of winter precipitation behaviour. Indeed, surface humidity fluxes can be reduced in winter by the sea ice cover, preventing the atmosphere from collecting moisture by evaporation. On fig. \ref{3D_struct}.d, similarly to the eastern part of the continent, precipitation rate and temperature are increasing along the slope to the ocean. Gradient changes in the precipitation curve at 73$^\circ$S and 78$^\circ$S are caused by topography.

\begin{figure*}[!h]
	\centering
	\includegraphics[width=1\linewidth]{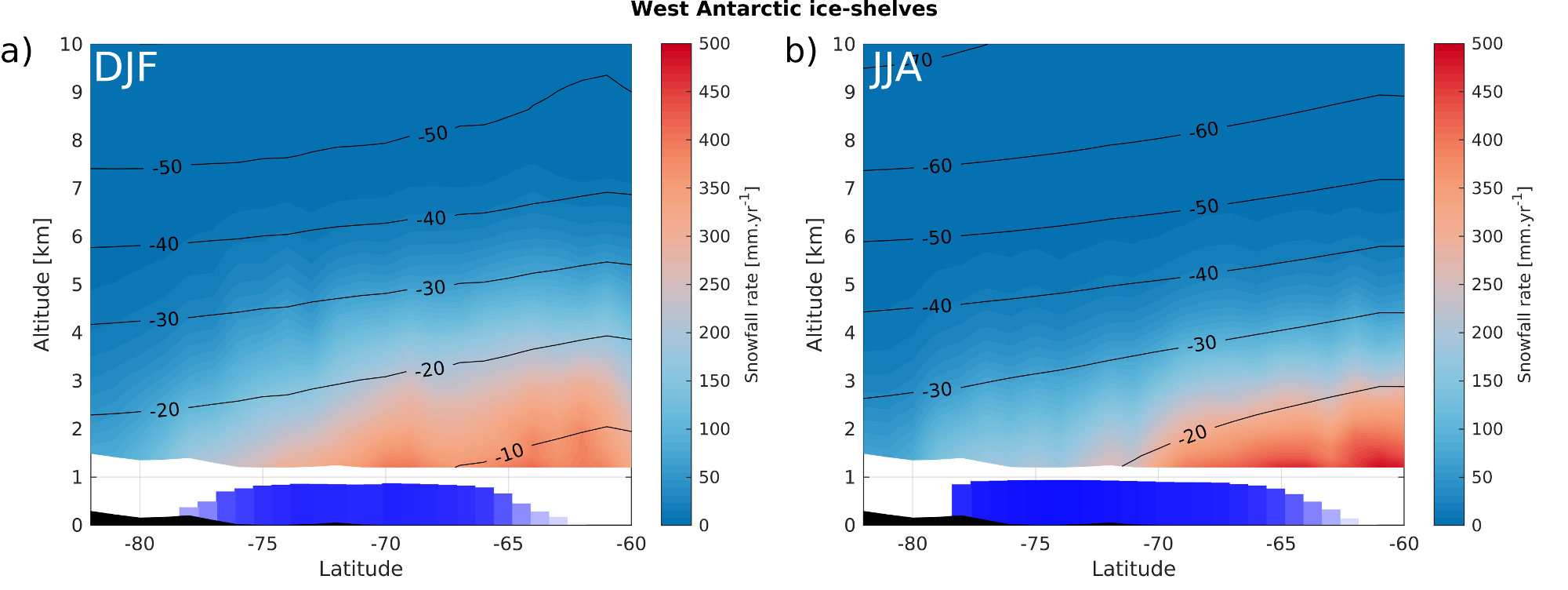}
	\caption{\label{3D_struct_IS}
		Zonal mean precipitation and its corresponding atmospheric temperature (contours) of the west Antarctic ice-shelves during the averaged summer and winter. The average period are \textbf{a)} December-January-February and \textbf{b)} June-July-August over the 2007-2010 period. The zonal averaged fractional sea-ice coverage obtained with ERA-Interim at a resolution of 0.75$^\circ$ is represented by the blue bars with a proportion ranging from 0 to 1.
	}
	\end{figure*}

Over the western ice-shelves on fig. \ref{3D_struct}.e, precipitation rate ranges from 100 to 500 mm.yr$^{-1}$ at the first level. This could be due to a lack of moisture input to the atmosphere due to the sea ice cover and ice-shelves. Indeed, there is a slight correlation between sea ice concentration and low level clouds \cite{schweiger2008relationships}. When crossing the ice-shelves at 75$^\circ$S, the precipitation increases quickly with latitude and altitude. Fig. \ref{3D_struct_IS} presents the seasonal variation in precipitation over this region. The reanalyses of ERA-Interim are used to present sea-ice cover fraction over the 2007--2010 period, which is shown in this figure using blue bars and ranging from 0 to 1 (0$\%$ to 100$\%$ of sea ice cover fraction). In winter (fig. \ref{3D_struct_IS}.b), the almost complete sea ice coverage shows a maximum precipitation at 60$^\circ$S which decreases rapidly above the ice-shelf following isotherms. In summer (fig. \ref{3D_struct_IS}.a), the average sea ice coverage is less extensive but still over 80$\%$. Also, there is higher averaged temperatures. Therefore, there is stronger precipitation rates closer to the coasts, with values ranging from 300 to 500 mm.yr$^{-1}$ in the vicinity of the coasts at 75$^\circ$S. Fig. \ref{3D_struct}.f shows a slow evolution of precipitation on ice-shelf with a very low standard deviation (less than 25$\%$ the value itself), while the temperature decreases linearly with latitude.
	
\begin{figure*}[!h]
	\centering
	\includegraphics[width=1\linewidth]{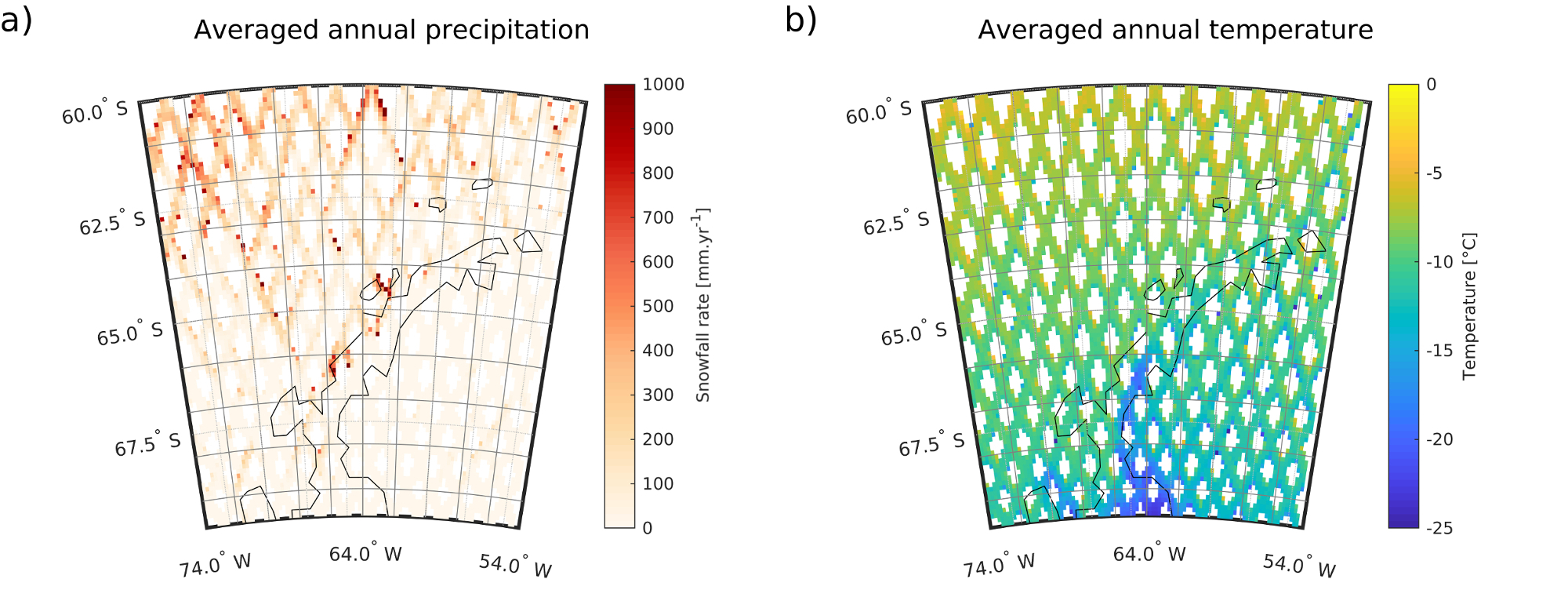}
	\caption{\label{Pen_subgrid}
	Averaged annual precipitation and temperature over the 2007--2010 period at 1200~m above ground level. The grid resolution is 0.1$^\circ$ in longitude by 0.1$^\circ$ in latitude. Precipitation is obtained from the 2C-SNOW-PROFILE product presented in section \ref{sect:data_methods} and temperature is obtained from the ECMWF-AUX product. The grey grid represents the resolution of the climatology presented in this study. CloudSat passes are not always located in exactly the same place. Thus the passes plotted on this figure with the red dots at the edge of the trackes may contain only a few strong precipitation events with a very low number of satellite overpasses ($<$10 passes). At this resolution and according to \citeA{souverijns2018cloudsatgrid}, there are more omissions than commissions of precipitation events by the CPR and few overflights. The average precipitation rate will thus could be very high.
	}
	\end{figure*}

The peninsula is a particular region of interest, with a mountain ridge extending across the circumpolar atmospheric circulation. Over this region, precipitation is mostly driven by orogenic effects. Fig. \ref{3D_struct}.h shows a high precipitation rate with a large standard deviation. The maximum snowfall rate is observed at the end of the peninsula with very high rates reaching 1000 mm.yr$^{-1}$ at 65$^\circ$S as seen on fig. \ref{3D_struct}.g. At 65$^\circ$S the precipitation reaches its maximum and there is an abrupt decrease in temperature. At this latitude, the topographic gradient of the peninsula is very intense. Over the peninsula, from 64$^\circ$S to 74$^\circ$S, precipitation rate is high and strongly affected by topography. It ranges from 500 to 700~mm.yr$^{-1}$ before decreasing from 75$^\circ$S to 82$^\circ$S with precipitation rates lower than 200~mm.yr$^{-1}$. The topographic obstacle at 75$^\circ$S with an average elevation of nearly 1000~m, but locally reaching 2500 m, results in two precipitation regimes, intense over the peninsula and low on the continental side. To better understand these results, we have performed a high-resolution climatology (see fig. \ref{Pen_subgrid}) of 0.1$^\circ$ in longitude by 0.1$^\circ$ in latitude on the first available level of CloudSat by averaging measured precipitations rates and corresponding temperatures in each cell as we did with a 2$^\circ$ in longitude by 1$^\circ$ in latitude grid for the whole continent (see section \ref{sect:data_methods}). This figure clearly shows the effect of topography on precipitation and temperature fields. Both temperature and precipitation are higher west of the peninsula, with very high precipitation rates reaching 1000~mm.yr$^{-1}$ observed along the west coast of the peninsula. On the east side, the temperature is lower due to the Larsen C ice-shelf. As seen in fig. \ref{Pen_subgrid}.b, the temperature above the peninsula is much lower than above the ocean with values ranging from -15 to -25$^\circ$C. This high-resolution view provides a better understanding of the precipitation and temperature variations observed in fig. \ref{3D_struct}.g and \ref{3D_struct}.h. Indeed, they show an average structure of warmer and wetter regions in the west of the peninsula with colder and drier regions in the east of the peninsula, thus increasing the variability of temperature and precipitation.
	
\begin{figure*} [!h]
	\centering
	\includegraphics[width=1\linewidth]{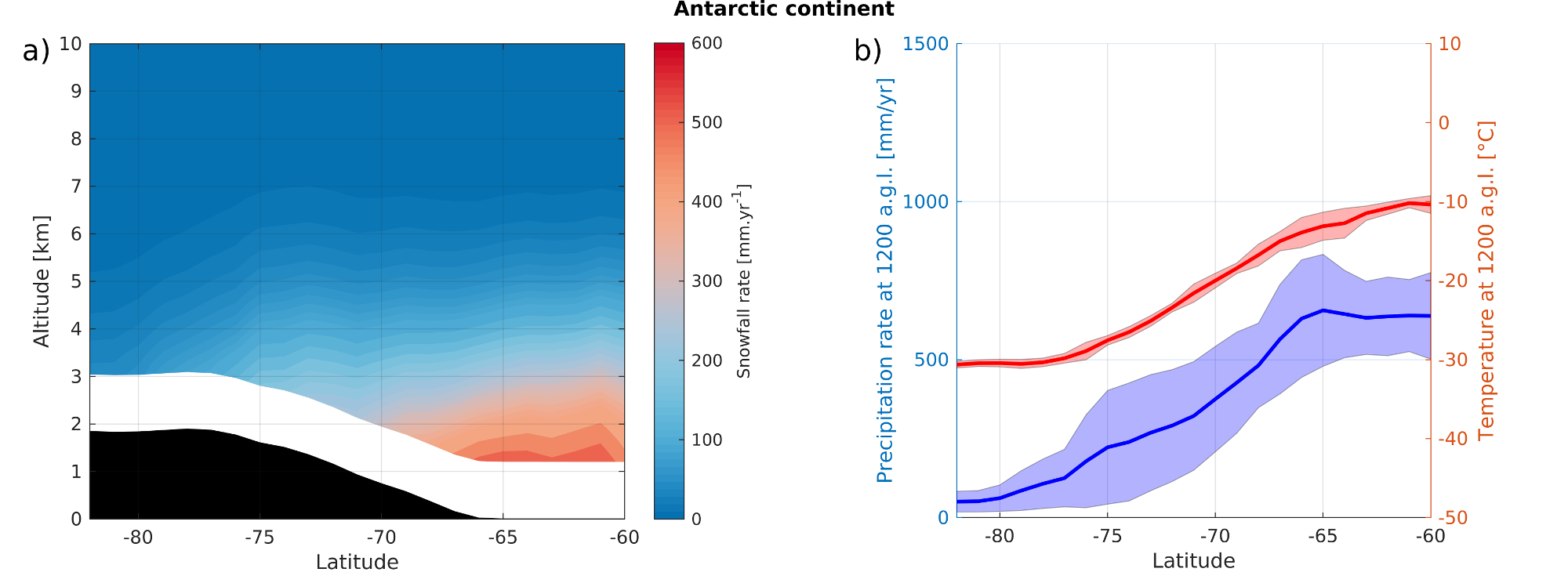}
	\caption{\label{General_3D}
	\textbf{a)} Latitudinally average of the precipitation structure and its corresponding atmospheric temperature (contours) for the entire continent.
	\textbf{b)} First CloudSat level of precipitation along the latitude (in blue) with its $\sigma$-spatial variation, and associated temperature (in red) with its $\sigma$-spatial standard deviation.
	}
	\end{figure*}

Fig. \ref{General_3D}.a summarizes the precipitation structure over the entire continent. It clearly shows that precipitation over ocean follows isotherms and thus evolves with temperature. Then precipitation isolines depart from isotherms over the topographic slope : precipitation rates are drastically decreasing when confronting the plateau. Indeed, the air masses rising along the topographical slope are now far from oceanic moisture sources, thus explaining the divergence of isotherms and precipitation contours. The average evolution of the topographical slope of Antarctica rises from sea level to 2000 metres above sea level from 66$^\circ$S to 78$^\circ$S. This topographical evolution is associated with an evolution of precipitation from 700~mm.yr$^{-1}$ to 150~mm.yr$^{-1}$. On the plateau, precipitation can reach values as low as 10~mm.yr$^{-1}$. Fig. \ref{General_3D}.b shows the evolution of the zonal mean precipitation rate at the fifth CloudSat bin. Over the plateau, precipitation rate is low but its standard deviation is relatively high, indicating that precipitation is not homogeneous, but brought by local events that transport moisture from the Southern Ocean. Along the topographic slope between coasts and plateau, precipitation and temperature are increasing, then precipitation reaches a maximum of 600~mm.yr$^{-1}$ at 66$^\circ$S.

\section{Precipitation distribution over the Antarctic continent}
\label{sect:scatterplots_precip}

In order to have a better representation of the precipitation over the ice-cap and to develop a new diagnostic tool for models, we looked at the distribution of the 2C-SNOW-PROFILE and ECMWF-AUX products raw data over the entire 3D dataset south of 60$^\circ$S. To do so, we have arranged these points in temperature bins of linear size and precipitation rate bins of logarithmic size. The study of the general structure of precipitation performed in section \ref{sect:3D_structure} showed that the main climatological differences are between the oceanic border regions and the high plateau. In this section we therefore keep distinguishing the peripheral precipitation from the precipitation over the plateau.

\subsection{Evolution of precipitation distributions with altitude}

\begin{figure*} [!h]
	\centering
	\includegraphics[width=1\linewidth]{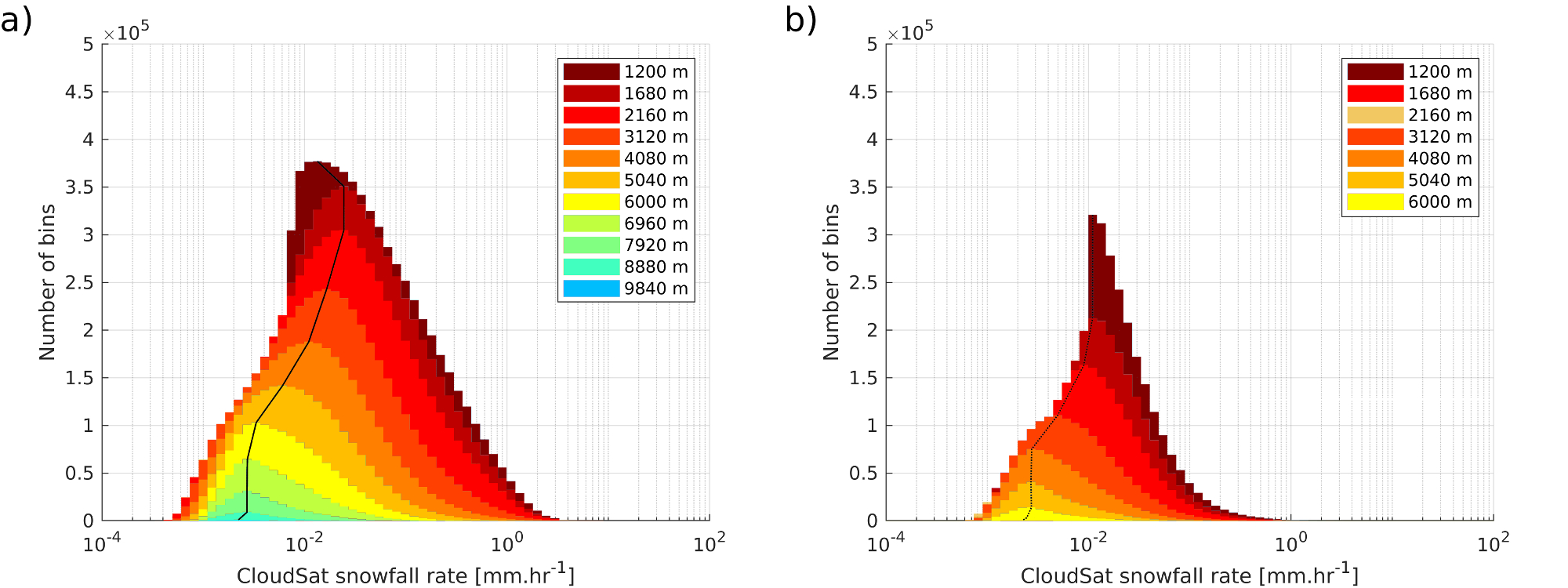}
	\caption{\label{Density_plot}
	Histogram plots of snowfall rates in mm.hr$^{-1}$ and vertical evolutions (\textbf{a)} for the peripheral areas ($<$2250~m) and \textbf{b)} for the plateau ($>$2250~m) at different altitudes above ground level. Black dashed lines are passing through maxima of each vertical bin distribution.
	}
	\end{figure*}

Each vertical CloudSat level above the surface has been studied separately. Fig. \ref{Density_plot} presents histograms of the distribution of precipitation rate observations evolving with altitude. In both areas, the precipitation distribution of the first level is truncated at 0.01 mm/hr, while for the other levels, there is no truncation. However, we observe precipitation rates below 0.01 mm/hr in the 2C-SNOW-PROFILE product. The Snow$\_$Retrieval$\_$Status values for these measurements are greater than 3, indicating that an error in the retrieval process could occur. It would therefore seem that ground clutter, in addition to generating excessive precipitation rates, does not allow sufficient confidence to be placed in small rates of near-surface precipitation. It is worth noting that we calculated the average precipitation rate at the second available level of CloudSat (at 1440 m.a.g.l.), by taking into account and by excluding the points recorded under 0.01 mm/hr to make sure that the ground clutter does not affect the averaged precipitation rate. This has no impact on the averaged snowfall rate at 1440 m.a.g.l. Doing so, we indeed verified that the missing observations do not affect the average precipitation rate at 1200 m.a.g.l., neither for peripheral regions nor on the plateau. 

We observe that the evolution of the number of observations and measured precipitation rates as a function of altitude above the coasts and plateau is similar, except for the first level under investigation. For coasts, distribution maxima at 1680 m.a.g.l. (7$^{th}$ bin) and 2160 m.a.g.l. (9$^{th}$ bin) do not vary in precipitation rate (0.0221 mm.hr$^{-1}$) but decrease from 14$\%$ in records numbers. It then reaches 0.0148 mm.hr$^{-1}$ at 3120 m.a.g.l. (13$^{th}$ bin), 0.0099 mm.hr$^{-1}$ at 4080 m.a.g.l. (17$^{th}$ bin) and 0.0055 mm.hr$^{-1}$ at 5040 m.a.g.l. (21$^{th}$ bin). Along these levels, there appears to be a consistency in the logarithmic decay of the peaks. Above 6000 m.a.g.l., the maximum precipitation rate stabilizes around a value of 0.0030 mm.hr$^{-1}$. Over the plateau, the position of the maximum in the lower vertical bins is at a lower precipitation rate than on the coasts. Apart from the truncated level, there is a decrease from 0.0099 mm.hr$^{-1}$ to 0.0081 mm.hr$^{-1}$ between 1680 m.a.g.l. (7$^{th}$ bin) and 2160 m.a.g.l. (9$^{th}$ bin). A thousand meters above at 4080 m.a.g.l. (17$^{th}$ bin) the peak is located at 0.0030 mm.hr$^{-1}$. Finally, at higher altitudes, the maximum remains in the same precipitation rate, centered on 0.0025 mm.hr$^{-1}$. For each region, the positions of the maxima at all altitudes are listed in Appendix A.

\subsection{Precipitation at saturation by forced lifting}

\begin{figure*} [!p]
	\centering
	\includegraphics[width=1\linewidth]{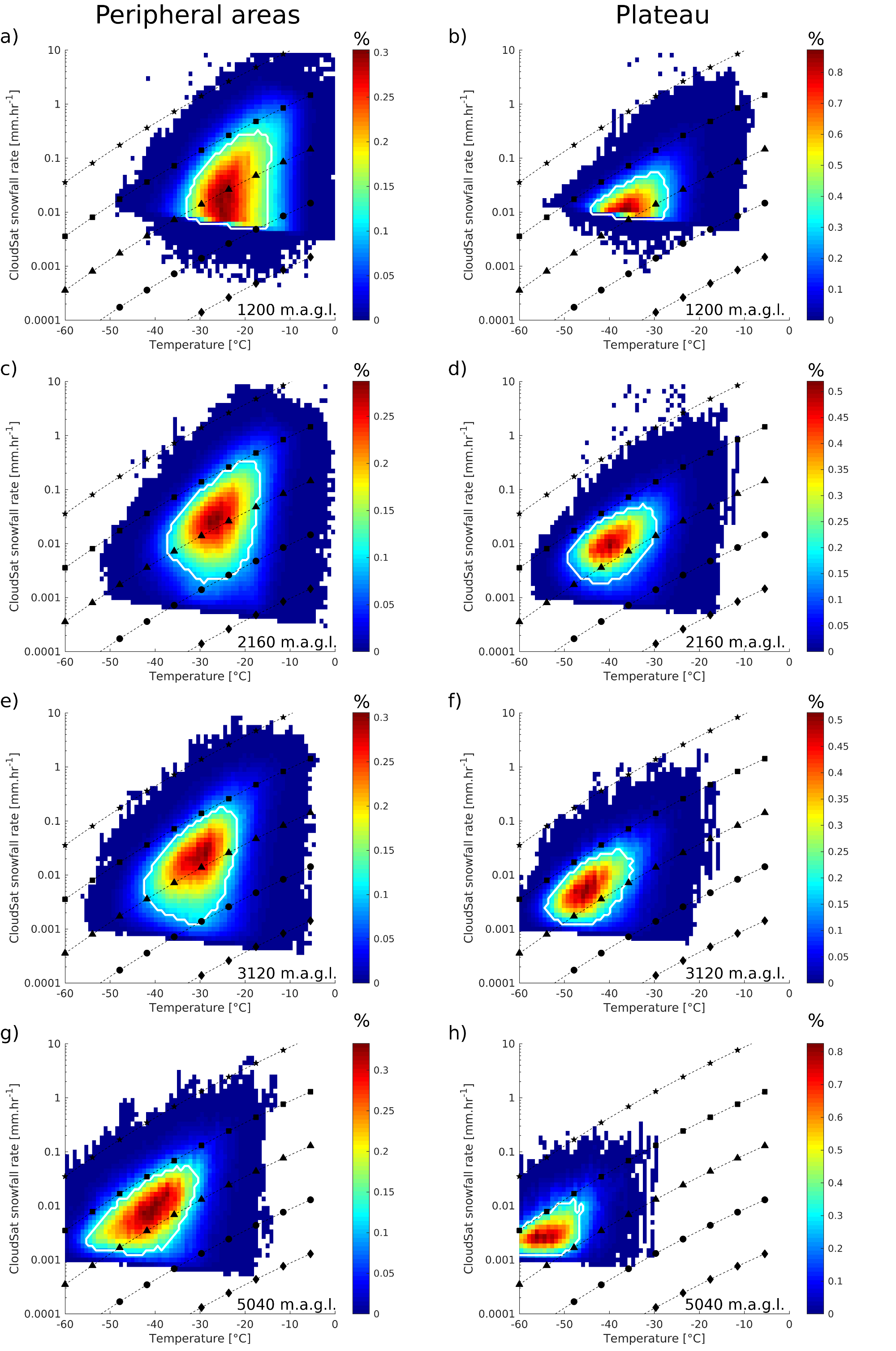}
	\caption{\label{Evolution}
	First column presents scatter plots of precipitation in mm/hr and temperature in $^\circ$C at different altitudes over the peripheral areas. Second column presents the same results over plateau area.
	The dashed black lines are the assumptions of theoretical precipitation rates calculated using equation \ref{eq:2_W_3} for vertical wind $w$ values of 0.0001~m.s$^{-1}$ (diamond markers), 0.001~m.s$^{-1}$ (circle markers), 0.01~m.s$^{-1}$ (triangle markers), 0.1~m.s$^{-1}$ (square markers) and 1~m.s$^{-1}$ (star markers).
	Colorbars are relative amounts of observations per CloudSat bin. White contours represent the $\sigma$ standard deviation of the distributions.
	}
	\end{figure*}

CloudSat observations are also presented on scatterplots in fig. \ref{Evolution} at different altitudes above ground level. The colorbar indicates the relative number of observations for a given precipitation rate and temperature. Over the peripheral regions, there is a significant spread both in precipitation rate and temperature. Over the plateau, the spread in precipitation is smaller and the distribution is centered at lower temperatures. In both regions, there seems to be a relationship between precipitation and temperature, and the spread in precipitation rate decreases when altitude increases. 

To further understand these distributions, we can derive an analytical relationship between precipitation and temperature by assuming that precipitation over the Antarctic ice cap is mostly controlled by the topographic lifting of oceanic air parcels that are initially close to saturation. We first assume that all the horizontal motion is converted into vertical motion. Then, assuming a mean horizontal wind velocity over a given slope, we can deduce the vertical speed of the oceanic air parcels. Then, this vertical wind speed can be used to lift air parcels along the moist adiabatic lapse rate. At each level, all the water vapor entering the parcel will be in excess of saturation and said to be equal to the precipitation mass flux. By integrating over the vertical direction, the resulting precipitation rate can be written as:

	\begin{equation}
	\label{eq:2_W_3}
	P_r = - \frac{w}{\rho_{water}}\int_{}^{z} \rho_{atm} \frac{L q_{sat}(T,p)}{R_{vap}T^2} \Gamma_{sat} dz
	\end{equation}

\noindent where $w$ is the vertical wind speed, $\rho_{water}$ is the water density and $\rho_{atm}$ is the air density, $L$ is the latent heat of sublimation, $q_{sat}$ is the humidity at saturation, $R_{vap}$ is the specific gas constant for wet air and $\Gamma_{sat}$ is the moist adiabatic lapse rate. The demonstration is in the \ref{(demo)}. This integral equation is resolved by using 240m $\Delta z$ corresponding to the CloudSat bin vertical dimensions. The precipitation $P_r$ expressed by equation \ref{eq:2_W_3} is given in water equivalent mm.hr$^{-1}$. It is represented by the black curves with markers on fig. \ref{Evolution}. The values used for $w$ are 0.0001~m.s$^{-1}$, 0.001~m.s$^{-1}$, 0.01~m.s$^{-1}$, 0.1~m.s$^{-1}$ and 1~m.s$^{-1}$. They are respectively represented in fig. \ref{Evolution} by black dashed lines with diamond, circle, triangle, square and star markers.  The rows are corresponding to CloudSat vertical bins above the surface, at respectively 1200, 2160, 3120 and 5040 m.a.g.l. which are corresponding to the 5$^{th}$, 9$^{th}$, 13$^{th}$ and 21$^{th}$ vertical bins above the surface. The white solid line $\sigma$ limit defining the standard deviation of the population distribution provides additional information on precipitation. In both cases, the spread in precipitation is greater when T increases and follows the trend with the triangles markers with a vertical velocity of 0.01 m/s. This vertical velocity $w$ can be assessed by a simple relationship between the slope $dz/dx$ and horizontal wind velocity. For example, based on fig. \ref{General_3D}, a horizontal wind $u~=~$5 m.s$^{-1}$ (this averaged wind velocity is in agreement with general circulation models, such as the IPSL-CM model in our case) advected over the average slope of the Antarctic ice cap, which extends over 10 degrees of latitude and 2000 m of vertical elevation gain, implies a vertical motion of $u.dz/dx = 0.01$ m.s$^{-1}$. A much lower vertical wind value implies an air advection along a gentler slope. A vertical speed of 0.1 m.s$^{-1}$ and above can be explained by sharp topographical obstacles, such as mountain ranges in West Antarctica and along the peninsula, or stronger large-scale winds. The higher the vertical velocity, the higher the precipitation rate.

At 1200 m.a.g.l. over peripheral areas on fig. \ref{Evolution}.a, most of the precipitation records are located between -32$^\circ$C and -14$^\circ$C, and between 0.006 mm.hr$^{-1}$ and 0.25 mm.hr$^{-1}$. The distribution is bounded between circle and square markers. When continuing to increase in altitude, the distribution follows the orientation of the analytical relationship. At the highest level considered in the fig. \ref{Evolution}.g, the spread is located between -55$^\circ$C and -31$^\circ$C in temperature and ranges from 0.001~mm.hr$^{-1}$ to 0.005~mm.hr$^{-1}$ in precipitation rate. Throughout the ascent along the CloudSat bins, the distribution evolves just over the line with the triangles markers, which means a vertical speed of about 0.02~m.s$^{-1}$. This is again consistent with an orographic precipitation on the margins of the ice-sheet. The table \ref{table_coasts} shows the precipitation and temperature locations of these distributions for levels from 5$^{th}$ to 25$^{th}$ CloudSat vertical bins. Above the plateau, on fig. \ref{Evolution}.b at 1200 m.a.g.l. the distribution is located between -43$^\circ$C and -27$^\circ$C. The lower boundary of the distribution is slightly above the theoretical triangles markers line. At this altitude, distribution is highly impacted by ground clutter. At a higher altitude of 2160 m.a.g.l. fig. \ref{Evolution}.d, this spread ranges from 0.002 mm.hr$^{-1}$ to 0.05 mm.hr$^{-1}$ with a temperature interval of -49$^\circ$C to -30$^\circ$C. As it evolves with altitude, the distribution continues to follow the line with the triangles markers, but only slightly above it. The table \ref{table_plateau} shows the precipitation and temperature locations of these distributions for levels from 5$^{th}$ to 25$^{th}$ CloudSat vertical bin.


	\begin{figure*} [!h]
	\centering
	\includegraphics[width=1\linewidth]{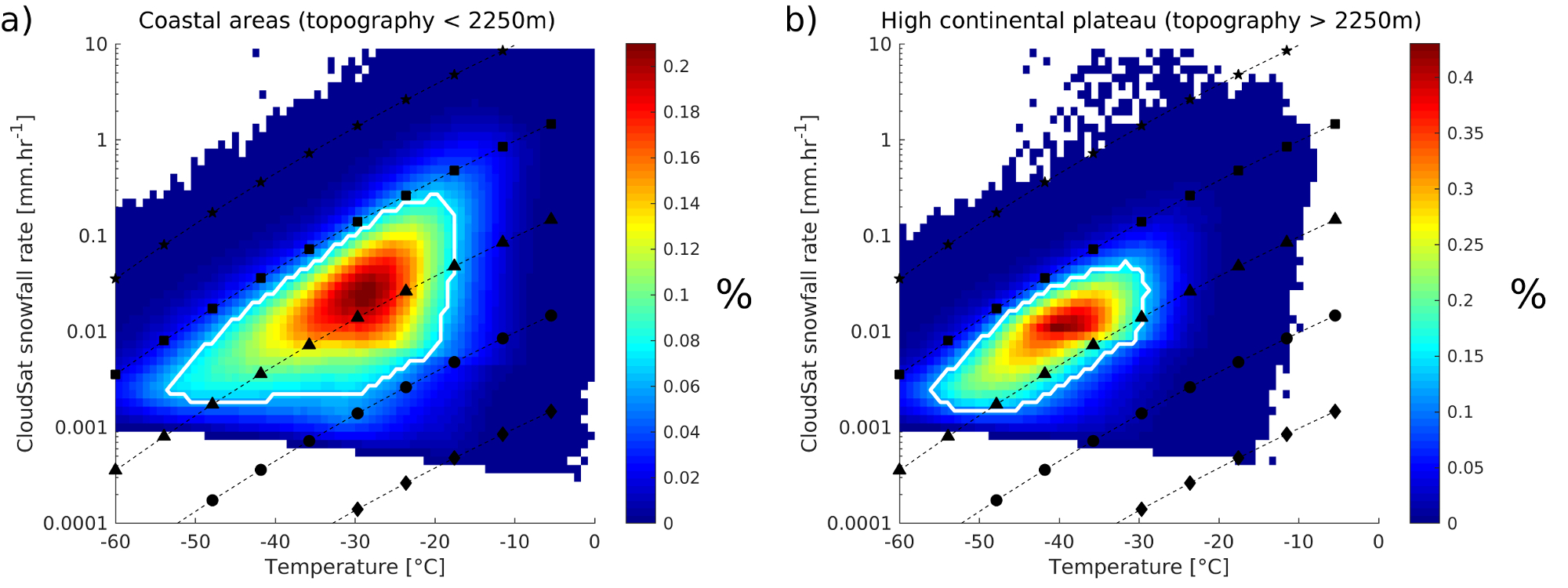}
	\caption{\label{Evolution_all}
	\textbf{a)} Scatter plots of precipitation in mm.hr$^{-1}$ and temperature in $^\circ$C at all altitudes over the peripheral areas. \textbf{a)} Same result over plateau area. The dashed black lines are the assumptions of theoretical precipitation rates calculated using equation \ref{eq:2_W_3} for vertical wind $w$ values of 0.0001~m.s$^{-1}$ (diamond markers), 0.001~m.s$^{-1}$ (circle markers), 0.01~m.s$^{-1}$ (triangle markers), 0.1~m.s$^{-1}$ (square markers) and 1~m.s$^{-1}$ (star markers). Colorbars are relative amounts of observations per CloudSat bin. White contours represent the $\sigma$ standard deviation of the distributions. Considered levels are 1200, 2160, 3120 and 5040 m.a.g.l.
	}
	\end{figure*}

\begin{figure*} [!h]
	\centering
	\includegraphics[width=0.5\linewidth]{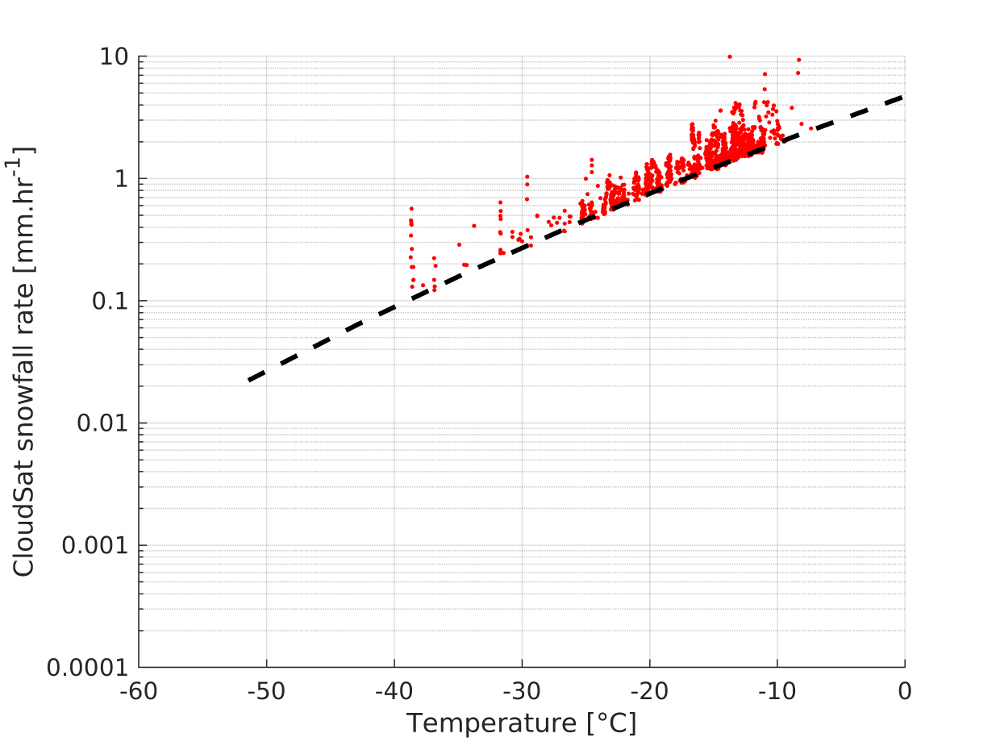}
	\caption{\label{Vent_fort}
	Observations on the plateau and coasts that exceed the resulting analytical relationship with a wind speed of 0.2 m.s$^{-1}$ (red scattered points). The dashed black line represents the analytical relationship with a vertical speed of 0.2~m.s$^{-1}$. There are 1251 records shown here.
	}
	\end{figure*}

\begin{figure*} [!h]
	\centering
	\includegraphics[width=1\linewidth]{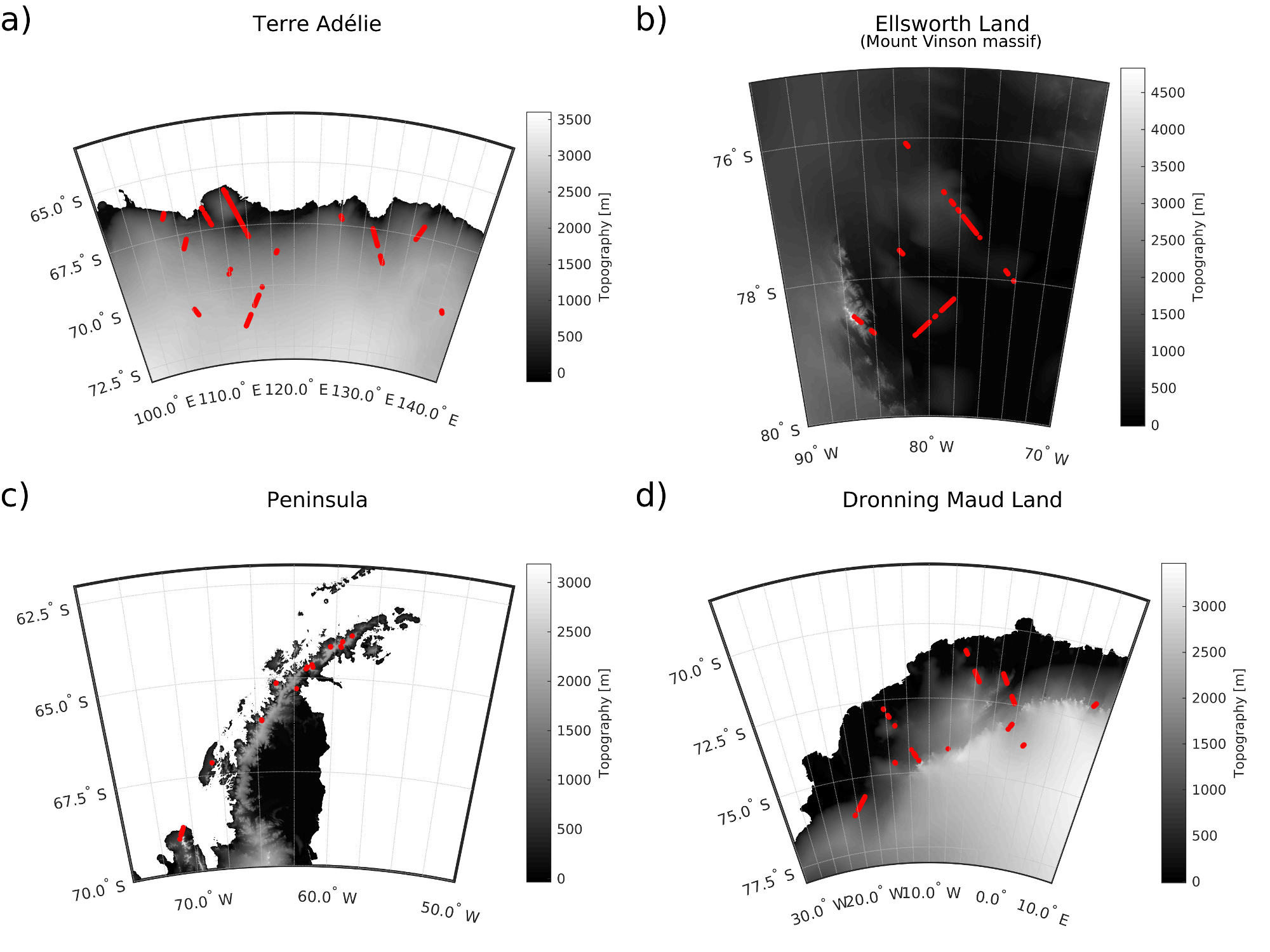}
	\caption{\label{Vent_fort_map}
	High resolution (200 m) topographic map \cite{greene2017antarctic,howat2019reference} over Terre Ad\'elie, Ellsworth Land, the peninsula and Dronning Maud Land. The red points correspond to the measurements greater than the 0.2 m.s$^{-1}$ vertical velocity precipitation hypothesis presented in fig. \ref{Vent_fort}.
	}
	\end{figure*}

Fig. \ref{Evolution_all} summarizes distribution of precipitation and temperature summed for all vertical levels. It shows that the distribution over peripheral regions reaches higher temperatures, so its precipitation dispersion is larger at these temperatures. Over peripheral regions, the density-plot is defined for large-scale vertical velocities $w$ ranging from 0.0001 (circle markers) to 0.1~m.s$^{-1}$ (square markers), while over the continental plateau, the distribution is bounded by triangle markers and square markers. Some observations reach very high precipitation rates, and are higher than the analytical relationship showed by square markers with a vertical velocity of 0.1 m.s$^{-1}$. This is either due to high large-scale winds analogous to extreme events or high slope analogous to local topographical obstacles along the peninsula and mountain ranges. We calculated a precipitation rate with a wind speed of 0.2 m.s$^{-1}$. This agrees either with a slope of 0.2$\%$, corresponding to the slope of the East Antarctic ice cap and a strong large-scale wind blowing at 100 m.s$^{-1}$ or with a mean large-scale wind of 5~m.s$^{-1}$ and a steep slope of 4$\%$. This covers many possible combinations of strong large scale winds and slopes ranging from the slope of the East Antarctic ice cap to much steeper slopes. Observations on the plateau and coasts that exceed the resulting analytical relationship with a wind speed of 0.2 m.s$^{-1}$ are showed on fig. \ref{Vent_fort}. In order to verify whether these measurements are made over areas of high topographic gradient, we have reported these measurements on a high-resolution topographic map \citeA{greene2017antarctic} and \citeA{howat2019reference} in figure \ref{Vent_fort_map}.

Fig. \ref{Vent_fort_map} shows a high-resolution (200 m) topographic map processed by \citeA{greene2017antarctic,howat2019reference} over 4 areas where these very high precipitation rates have been recorded by CloudSat. These measurements suggest a vertical velocity of advection at a large scale greater than 0.1 m.s$^{-1}$, possible only in regions with a high topographic gradient. The slope of the region of Terre Ad\'elie showed on fig. \ref{Vent_fort_map}.a respects the topographic criterion causing strong vertical winds. Fig. \ref{Vent_fort_map}.b presenting Ellsworth Land and Mount Vinson massif satisfies the criterion of a high topographic gradient. Indeed, the Mount Vinson massif is a mountain range and the highest point in Antarctica. The peninsula on fig. \ref{Vent_fort_map}.c, as previously discussed, is a topographic barrier that crosses and obstructs areas with major circumpolar air currents. And the fourth region on fig. \ref{Vent_fort_map}.d, Dronning Maud, is also respecting the topographic criterion causing strong vertical winds with the presence of a mountainous barrier a few hundred kilometres from the coast. On the four cases, the red dots corresponding to high precipitation measurements follow each others, they correspond to the same satellite track at a given time. These are rare occasional events of massive precipitation that contribute significantly to the accumulation of snow on the continent.
	
\begin{figure*} [!h]
	\centering
	\includegraphics[width=1\linewidth]{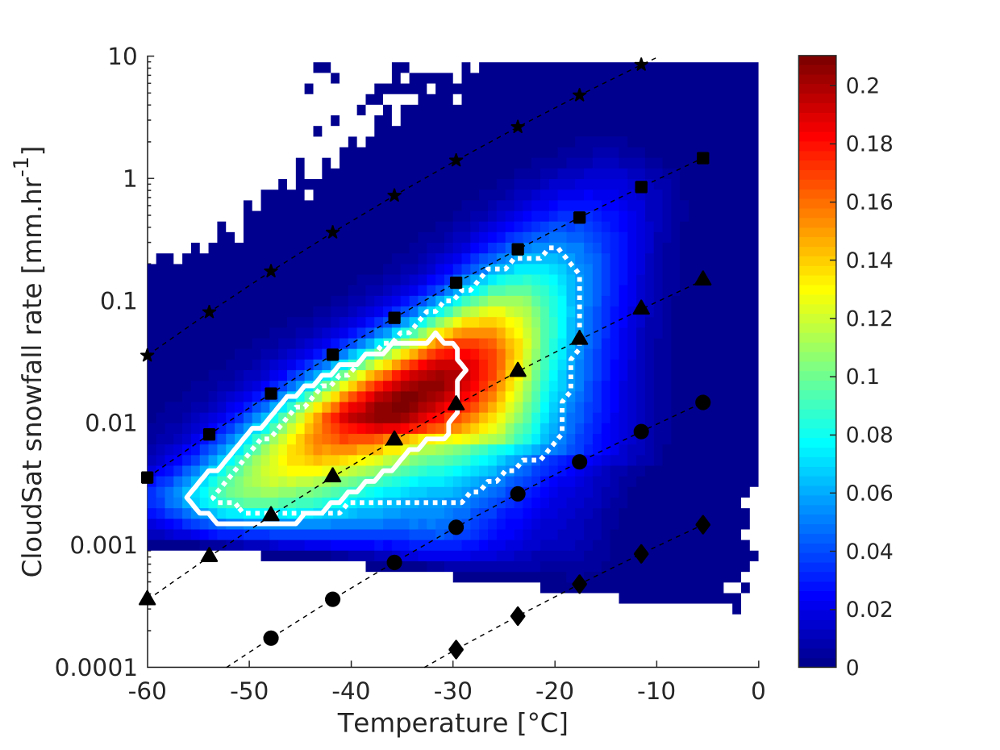}
	\caption{\label{Fig_finale}
	Scatter plots of precipitation in mm.hr$^{-1}$ and temperature in $^\circ$C at all altitudes over the Antarctic continent. The colorbar indicates the relative number of observations for a given precipitation rate and temperature. The dashed black lines are the assumptions of theoretical precipitation rates calculated with the moisture convergence equation for vertical wind $w$ values of 0.0001~m.s$^{-1}$ (diamond markers), 0.001~m.s$^{-1}$ (circle markers), 0.01~m.s$^{-1}$ (triangle markers) 0.1~m.s$^{-1}$ (square markers) and 1~m.s$^{-1}$ (star markers). Dotted white line represents the 1-$\sigma$ distribution of precipitation over the peripheral areas and solid white line represents the 1-$\sigma$ distribution of precipitation over the plateau.
	}
	\end{figure*}

Fig. \ref{Fig_finale} summarizes the behaviour of precipitation over the Antarctic continent. Precipitation over the peripheral areas occurs with higher temperature by forced lifting of air masses along the topographic slope. The different topographical slopes as well as the strength of the large-scale horizontal wind thus generates different vertical wind velocities. According to equation \ref{eq:2_W_3}, the variability of these winds consequently generates a wide spread in precipitation rates, identified by the white dashed line contours. Then over the plateau, the available water quantities, temperatures, low slopes and low large-scale horizontal winds cause precipitation to evolve following Clausius-Clapeyron relation with a small spread in precipitation rate. This is represented on fig. \ref{Fig_finale} by the white solid contours.

\section{Conclusion}
\label{sect:conclusion}

Precipitation is mostly considered as a surface variable, climatologies typically only reporting the 2D horizontal distribution at the surface. The CloudSat radar dataset now allows to explore the 3D structure of precipitation, and this study provides insight into the origins of precipitation and its evolution along the vertical dimension over Antarctica. The 2C-SNOW-PROFILE product has been explored on the three spatial dimensions and the temporal dimension from 2007 to 2010 over the Antarctic continent. This 3D-dataset is computed following \citeA{palerme2014much,palerme2018groundclutter} with a horizontal resolution of 1$^\circ$ of latitude by 2$^\circ$ of longitude grid and a 240~m vertical resolution in order to optimally represent the Southern polar climate. It has been studied over several different regions, geographically divided into  four regions (East continent, West continent, West ice-shelves and Peninsula), and topographically divided into two regions: the peripheral areas (where topography $<$ 2250 m) and the Antarctic plateau (where topography $>$ 2250 m). The distinction between peripheral areas and continental plateau revealed that most of the precipitation is located over the peripheral areas with low relative seasonal variability, in contrast with the plateau where it is relatively high. The study of the four geographical regions revealed many differences between the Peninsula and the western ice-shelves as compared to the continental eastern and western parts. Indeed, the Peninsula is an area where precipitation is strong and mainly driven by local orography and snowfall rates are low over the ice-shelves, while precipitation is increasing along the slope to the ocean over continental eastern and western regions. There also seems to be a strong dichotomy in the behaviour of precipitation over the ice-shelves. This could be related to the sea ice cover, which is able to stratify the atmosphere by cooling it, thus limiting the mixing and thus precipitation.

The pre-gridded CloudSat product is then statistically studied in order to evaluate the distribution of snowfall rates and associated temperatures over the peripheral areas and the plateau at each CloudSat vertical level. This shows that precipitation, whatever the altitude, is mainly driven by large-scale convergence of moist fluxes over the topography. This has been confirmed by comparing the observed precipitation/temperature distributions with an analytical relationship of precipitation as a function of temperature and a vertical advection velocity that is directly dependent on large scale horizontal wind strength and slope. Over the peripheral regions, the dispersion of precipitation rates as a function of temperature is large. This may mean that both topographical slopes and vertical winds are highly diversified. Over the continental plateau, the dispersion is reduced and results from gentler slopes than in peripheral regions, as well as slower vertical uplifts. Thus, the precipitation dispersion at a given temperature of these distributions is justified by varying degrees of horizontal advection on variable slopes. The averaged observed precipitation distribution follows the analytical curve corresponding to a vertical wind of 0.01 m.s$^{-1}$, which is typical of the values found in general circulation models, for example in our case the IPSL-CM. This new study of the CloudSat precipitation product provides new and innovative tools to evaluate climate models with a three-dimensional view of the atmospheric structure of precipitation. On the one hand, the use of average continental precipitation profiles and zonal averages are diagnostic tools that are easily comparable to models. On the other hand, precipitation rate and temperature scatter-plots require high-frequency model outputs, but lead to a powerful way of evaluating transport, thermodynamics and microphysical processes in further detail.

\appendix

	\section{Location of the maximum precipitation rate as a function of altitude}
	
	Selected bin values for precipitation rate and temperature as presented in figures \ref{Density_plot}, \ref{Evolution_all}, \ref{Vent_fort} and \ref{Evolution}.
	
	\noindent Temperature: \\
	\noindent[-60.0000	-59.1429	-58.2857	-57.4286	-56.5714	-55.7143	-54.8571	-54.0000	-53.1429	-52.2857	-51.4286	-50.5714	-49.7143	-48.8571	-48.0000	-47.1429	-46.2857	-45.4286	-44.5714	-43.7143	-42.8571	-42.0000	-41.1429	-40.2857	-39.4286	-38.5714	-37.7143	-36.8571	-36.0000	-35.1429	-34.2857	-33.4286	-32.5714	-31.7143	-30.8571	-30.0000	-29.1429	-28.2857	-27.4286	-26.5714	-25.7143	-24.8571	-24.0000	-23.1426	-22.2857	-21.4286	-20.5714	-19.7143	-18.8571	-18.0000	-17.1429	-16.2857	-15.4286	-14.5714	-13.7143	-12.8571	-12.0000	-11.1429	-10.2857	-9.4286	-8.5714	-7.7143	-6.8571	-6.0000	-5.1429	-4.2857	-3.4286	-2.5714	-1.7143	-0.8571	0.0000]
	
	\noindent Precipitation rate:
	\noindent[0.0001	0.0001	0.0001	0.0002	0.0002	0.0003	0.0003	0.0004	0.0005	0.0006	0.0007	0.0009	0.0011	0.0013	0.0016	0.0020	0.0025	0.0030	0.0037	0.0045	0.0055	0.0067	0.0081	0.0099	0.0122	0.0148	0.0181	0.0221	0.0270	0.0330	0.0403	0.0493	0.0602	0.0735	0.0898	0.1097	0.1340	0.1636	0.1998	0.2441	0.2981	0.3641	0.4447	0.5432	0.6634	0.8103	0.9897	1.2088	1.4765	1.8034	2.2026	2.6903	3.2850	4.0139	4.9021	5.9874	7.3130	8.9321]
	
	The number of decimals for these bins after the decimal point was chosen according to the sensitivity of the CPR measurements.
	
	\begin{table}[!p]
	\centering
	\caption{\label{table_pic} Location of the maximum precipitation rate as a function of altitude, as presented in figure \ref{Density_plot}. Values for all CloudSat bins ranging from 1200 m.a.g.l. (5$^{th}$ bin) to 6000 m.a.g.l. (25$^{th}$ bin) above the surface are presented. The format is as follows: 'precipitation rate in mm.hr$^{-1}$ -- number of observations'.
	}
	\begin{tabular}{lrr}
	\hline
	Altitude above ground level & Coastal areas & Continental plateau \\
	\hline

	1200 m		& 0.0122 -- 377278 & 0.0099 -- 321044 \\
	1440 m		& 0.0181 -- 377234 & 0.0122 -- 260956 \\
	1680 m		& 0.0221 -- 351080 & 0.0099 -- 212493 \\
	1920 m		& 0.0221 -- 325865 & 0.0099 -- 184575 \\
	2160 m		& 0.0221 -- 305071 & 0.0081 -- 163445 \\
	2400 m		& 0.0181 -- 288393 & 0.0067 -- 146641 \\
	2640 m		& 0.0181 -- 272086 & 0.0055 -- 132604 \\
	2880 m		& 0.0181 -- 257160 & 0.0055 -- 121014 \\
	3120 m		& 0.0148 -- 243191 & 0.0045 -- 110872 \\
	3360 m		& 0.0121 -- 229763 & 0.0037 -- 100530 \\
	3600 m		& 0.0121 -- 216356 & 0.0030 -- 91289 \\
	3840 m		& 0.0099 -- 201847 & 0.0030 -- 83873 \\
	4080 m		& 0.0099 -- 187903 & 0.0030 -- 75265 \\
	4320 m		& 0.0081 -- 175297 & 0.0025 -- 66611 \\
	4560 m		& 0.0081 -- 164391 & 0.0025 -- 57850 \\
	4800 m		& 0.0067 -- 153230 & 0.0025 -- 49477 \\
	5040 m		& 0.0055 -- 142044 & 0.0025 -- 40760 \\
	5280 m		& 0.0045 -- 131296 & 0.0025 -- 32798 \\
	5520 m		& 0.0037 -- 121678 & 0.0025 -- 25271 \\
	5760 m		& 0.0030 -- 111769 & 0.0025 -- 19258 \\
	6000 m		& 0.0030 -- 103270 & 0.0025 -- 14080 \\	
	\hline
	\end{tabular}
	\end{table}
	
	\begin{table}[!p]
	\centering
	\caption{\label{table_coasts} Location of the $\sigma$ deviation of the maximum precipitation rate as a function of altitude for the peripheral areas. Values for all CloudSat bins ranging from 1200 m.a.g.l. (5$^{th}$ bin) to 6000 m.a.g.l. (25$^{th}$ bin) above the surface are presented. The format is as follows: 'maximum observation concentration'[extrema extrema]. The bold values indicate that the maximum distribution is overlaid with the 1-$\sigma$ distribution boundary.
	}
	\begin{tabular}{lrr}
	\hline
	Altitude above ground level & Precipitation rate in mm.hr$^-1$ & Temperature in $^\circ$C \\
	\hline

	1200 m		& 0.0148 [0.0055 -- 0.2981] & -25.7143 [-32.5714 -- -14.5714] \\
	1440 m		& 0.0181 [0.0055 -- 0.2981] & -25.7143 [-33.4286 -- -14.5714] \\
	1680 m		& 0.0270 [0.0037 -- 0.3641] & -25.7143 [-35.1429 -- -14.5714] \\
	1920 m		& 0.0270 [0.0030 -- 0.2981] & -26.5714 [-36.0000 -- -16.2857] \\
	2160 m		& 0.0221 [0.0020 -- 0.2981] & -28.2857 [-17.1429 -- -36.8571] \\
	2400 m		& 0.0221 [0.0020 -- 0.2441] & -30.0000 [-38.5714 -- -18.0000] \\
	2640 m		& 0.0270 [0.0013 -- 0.1998] & -28.2857 [-40.2857 -- -18.8571] \\
	2880 m		& 0.0270 [0.0013 -- 0.1636] & -28.2857 [-41.1429 -- -20.5714] \\
	3120 m		& 0.0221 [0.0013 -- 0.1636] & -29.1429 [-42.0000 -- -22.2857] \\
	3360 m		& 0.0221 [0.0013 -- 0.1339] & -30.0000 [-43.7143 -- -23.1429] \\
	3600 m		& 0.0181 [0.0013 -- 0.1339] & -31.7143 [-45.4286 -- -24.0000] \\
	3840 m		& 0.0122 [0.0016 -- 0.1097] & -35.1429 [-48.0000 -- -24.8571] \\
	4080 m		& 0.0148 [0.0016 -- 0.0898] & -36.0000 [-48.8571 -- -26.5714] \\
	4320 m		& 0.0099 [0.0013 -- 0.0735] & -38.5714 [-49.7143 -- -28.2857] \\
	4560 m		& 0.0099 [0.0013 -- 0.0602] & -37.7143 [-50.5714 -- -29.1429] \\
	4800 m		& 0.0067 [0.0013 -- 0.0602] & \textbf{-41.1429} [\textbf{-41.1429} -- -30.0000] \\
	5040 m		& 0.0099 [0.0013 -- 0.0493] & -40.2857 [-54.8571 -- -30.8571] \\
	5280 m		& 0.0055 [0.0013 -- 0.0404] & -43.7143 [-55.7143 -- -33.4286] \\
	5520 m		& 0.0055 [0.0013 -- 0.0330] & -45.4286 [-56.5714 -- -35.1429] \\
	5760 m		& 0.0037 [0.0013 -- 0.0270] & -48.0000 [-56.5714 -- -35.1429] \\
	6000 m		& 0.0030 [0.0013 -- 0.0270] & -48.0000 [-57.4286 -- -36.0000] \\	
	\hline
	\end{tabular}
	\end{table}
		
	\begin{table}[!p]
	\centering
	\caption{\label{table_plateau} Location of the $\sigma$ deviation of the maximum precipitation rate as a function of altitude for the plateau. Values for all CloudSat bins ranging from 1200 m.a.g.l. (5$^{th}$ bin) to 6000 m.a.g.l. (25$^{th}$ bin) above the surface are presented. The format is as follows: 'maximum observation concentration'[extrema extrema].
	}
	\begin{tabular}{lrr}
	\hline
	Altitude above ground level & Precipitation rate in mm.hr$^-1$ & Temperature in $^\circ$C \\
	\hline

	1200 m		& 0.0099 [0.0081 -- 0.0493] & -36.8571 [-43.7143 -- -27.4286] \\
	1440 m		& 0.0122 [0.0055 -- 0.0602] & -37.7143 [-45.4286 -- -27.4286] \\
	1680 m		& 0.0099 [0.0037 -- 0.0602] & -38.5714 [-46.2857 -- -28.2857] \\
	1920 m		& 0.0099 [0.0030 -- 0.0493] & -38.5714 [-47.1429 -- -28.2857] \\
	2160 m		& 0.0081 [0.0020 -- 0.0493] & -40.2857 [-48.8571 -- -30.0000] \\
	2400 m		& 0.0099 [0.0016 -- 0.0403] & -40.2857 [-49.7143 -- -31.7143] \\
	2640 m		& 0.0067 [0.0013 -- 0.0330] & -42.0000 [-50.5714 -- -32.5714] \\
	2880 m		& 0.0055 [0.0013 -- 0.0330] & -43.7143 [-52.2857 -- -34.2857] \\
	3120 m		& 0.0055 [0.0013 -- 0.0270] & -45.4286 [-54.0000 -- -35.1429] \\
	3360 m		& 0.0037 [0.0013 -- 0.0221] & -48.0000 [-54.8571 -- -36.8571] \\
	3600 m		& 0.0037 [0.0013 -- 0.0221] & -48.8571 [-55.7143 -- -38.5714] \\
	3840 m		& 0.0030 [0.0016 -- 0.0181] & -49.7143 [-56.5714 -- -39.4286] \\
	4080 m		& 0.0030 [0.0011 -- 0.0181] & -51.4286 [-58.2857 -- -41.1429] \\
	4320 m		& 0.0025 [0.0011 -- 0.0148] & -53.1429 [-59.1429 -- -42.8571] \\
	4560 m		& 0.0030 [0.0011 -- 0.0122] & -53.1429 [-50.5714 -- -29.1429] \\
	4800 m		& 0.0025 [0.0011 -- 0.0122] & -54.8571 [-60.0000 -- -45.4286] \\
	5040 m		& 0.0025 [0.0013 -- 0.0122] & -56.5714 [-60.0000 -- -46.2857] \\
	5280 m		& 0.0025 [0.0013 -- 0.0099] & -56.5714 [-60.0000 -- -46.2857] \\
	5520 m		& 0.0025 [0.0011 -- 0.0099] & -59.1429 [-60.0000 -- -48.0000] \\
	5760 m		& 0.0025 [0.0011 -- 0.0099] & -59.1429 [-60.0000 -- -49.7143] \\
	6000 m		& 0.0025 [0.0011 -- 0.0081] & -55.7143 [-60.0000 -- -49.7143] \\	
	\hline
	\end{tabular}
	\end{table}

\section{Precipitation at saturation by forced lifting demonstation}
\label{(demo)}
All the air advected into the pole would condense by forced lifting due to the topographical landmass of the continent. We assume a purely vertical motion initiated from a horizontal advection at the origin of large-scale precipitation. The vertical motion of the air mass implies a cooling and a variation in the saturation vapour pressure, and therefore in the precipitation rate. We use moisture flow convergence equations and test several vertical wind values in order to explain precipitation observations. This is based on the net moisture balance in a steady state saturated atmosphere and can be derived from:

	\begin{equation}
	\label{eq:2_W_1}
	\frac{Dq}{Dt} = S
	\end{equation}

\noindent where $D/Dt$ is the lagrangian derivative:

	\begin{equation}
	\frac{D}{Dt} = \frac{\partial}{\partial t} + u \frac{\partial}{\partial x} + v \frac{\partial}{\partial y} + w \frac{\partial}{\partial z}
	\end{equation}

\noindent with $u$, $v$ and $w$, representing the standard three-dimensional wind components, and $q$ is the specific humidity. $S$ represents the budget of water vapor, which is defined by the difference between sources and sinks of water vapor following air parcel motion. $S$ typically takes the form $Ev-Pr$, where $Ev$ is the evaporation rate into the air parcel and $Pr$ is the precipitation rate.

	\begin{equation}
	\frac{\partial q}{ \partial t} + u \frac{\partial q}{\partial x}  + v \frac{\partial q}{\partial y} + w \frac{\partial q}{\partial z} = Ev-Pr
	\end{equation}
	
\noindent If we assume that the lifted parcels are brought to saturation, $Ev~=~0$ kg.kg$^{-1}$.s$^{-1}$, $q=q_{sat}(T,p)$ and in steady state, ${\partial q}/{\partial t}~=~0$~kg.kg$^{-1}$.s$^{-1}$ so:

	\begin{equation}
	\label{eq:2_W_2}
	\vec{u}.\vec{\nabla}q_{sat}(T,p) = - Pr
	\end{equation}

\noindent Assuming a purely vertical motion, we finally reach the form:

	\begin{equation}
	w\frac{\partial q_{sat}(T,p)}{\partial z} = - Pr
	\end{equation}

\noindent We can therefore assess the precipitation rate by integrating a vertical flux over the column observed by CloudSat at a given constant $w$:

	\begin{equation}
	P_{r} = - w\int_{}^{z}\rho_{atm}\frac{\partial q_{sat}(T,p)}{\partial z}dz
	\end{equation}

\noindent The term ${\partial q_{sat}(T,p)}/{\partial z}$ can be decomposed and then described from Clausius Clapeyron's equation as follow:

	\begin{equation}
	\frac{de_{sat}}{e_{sat}} = \frac{L}{R_{vap}T^2}dT
	\end{equation}

\noindent with $q_{sat}(T,p)~=~0.622~e_{sat}/p$. At a constant pressure:

	\begin{equation}
	\frac{\partial q_{sat}}{\partial z} = \frac{\partial q_{sat}(T,p)}{\partial T}.\frac{\partial T}{\partial z}=\frac{Lq_{sat}(T,p)}{R_{vap}T^2}\frac{\partial T}{\partial z}
	\end{equation}

\noindent with the latent heat of sublimation $L$ and the specific gas constant for wet air $R_{vap}$. Since we assume that we are at saturation, $\partial T$~/~$\partial z$ is the moist adiabatic lapse rate $\Gamma_{sat}$ with a value of -6.5~K.km$^{-1}$. The definitive form of the precipitation equation is:
	
	\begin{equation}
	P_{r} = - w\int_{}^{z} \rho_{atm} \frac{L q_{sat}(T,p)}{R_{vap}T^2} \Gamma_{sat} dz
	\end{equation}

\acknowledgments
This work was supported by the French National Research Agency (Grant number : ANR-15-CE01-0003). CloudSat data is freely available via the CloudSat Data Processing Center (http://www.cloudsat.cira.colostate.edu/). The authors thank Karine Marquois and the IT department of the Laboratoire de M\'et\'eorologie Dynamique for the informatics support to generate the 3D climatology. The authors thank Anna-Lea Albright, Margaux Valls, Luca Montabone and M\'elanie Thiriet for the proofreading.
\newpage

\bibliography{Biblio_These}

\end{document}